\documentclass[10pt]{article}

\usepackage{array}
\usepackage{color} 
\usepackage{mathrsfs}
\usepackage{graphicx}
\usepackage{amsmath}
\usepackage{bbm}
\usepackage{amsfonts}
\usepackage{amssymb}
\usepackage{amsbsy}
\usepackage[T1]{fontenc}
\usepackage{theorem}
\usepackage{stmaryrd}
\usepackage{tipa}
\usepackage{textcomp}
\usepackage{subfigure}
\usepackage{epsfig}
\usepackage[numbers]{natbib}
\usepackage{lmodern}
\usepackage{framed}
\usepackage[subfigure]{tocloft}
\usepackage{subeqnarray}
\usepackage{alphalph}
\usepackage[refpage,cfg,prefix]{nomencl}
\usepackage{multirow}
\usepackage{xifthen}
\usepackage{enumerate}
\usepackage{color}
\usepackage{wasysym}
\usepackage{bibentry}
\usepackage{todonotes}
\nobibliography*
\usepackage{verbatim}
\usepackage{booktabs}
\usepackage[hyperindex,breaklinks]{hyperref}
\hypersetup{
colorlinks=true,
urlcolor=blue,
linkcolor=black,
citecolor=black,
}
\usepackage{breakurl}
\usepackage{url}
\usepackage{setspace}
\usepackage{xr}
\usepackage{wrapfig}
\usepackage{mathtools}
\usepackage{dsfont}
\usepackage[toc,page]{appendix}

\makenomenclature

\usepackage[labelfont=bf,labelsep=period,justification=raggedright]{caption}

\makeatletter
\renewcommand{\@biblabel}[1]{\quad#1.}
\makeatother

\allowdisplaybreaks

\date{}

\newcommand{\bs}[1]{\boldsymbol{#1}}

\newcommand{\partder}[2]{\frac{\partial #1}{\partial #2}}
\newcommand{\ordder}[2]{\frac{\text{d} #1}{\text{d} #2}}

\newcommand{\secpartder}[2]{\frac{\partial^2 #1}{\partial #2 ^2}}

\newcommand{\set}[2]{\left\{#1:#2\right\}}

\newcommand{\listofalgorithms}{\textbf{\Huge{List of Algorithms}}}
\newlistof{Algorithm}{exp}{\listofalgorithms}
\newcounter{instructioncounter}
 
\newenvironment{Algorithm}[2][]
{\refstepcounter{Algorithm}
\begin{framed}\addcontentsline{exp}{Algorithm}{\protect\numberline{\theAlgorithm} #1}\par\begin{center}\textbf{Algorithm \theAlgorithm : #2}\end{center}\begin{list}
{\bf{(\arabic{Algorithm}\alph{instructioncounter}})}{\usecounter{instructioncounter}}}{\end{list}\end{framed}}

\graphicspath{{./Images/}}

\begin{document}

\begin{flushleft}
{\Large
\textbf{Unbiased on-lattice domain growth}
}
\\
Cameron A. Smith$^{1,\ast}$, 
C\'{e}cile Mailler$^{2}$,
Christian A. Yates$^{1}$
\\
$^1$\textbf{Centre for Mathematical Biology, Department of Mathematical Sciences, University of Bath, Claverton Down, Bath, BA2 7AY, United Kingdom}\\
$^2$\textbf{Probability Laboratory, Department of Mathematical Sciences, University of Bath, Claverton Down, Bath, BA2 7AY, United Kingdom\\
$^\ast$E-mail: c.smith3@bath.ac.uk}
\end{flushleft}


\begin{abstract}
Domain growth is a key process in many areas of biology, including embryonic development, the growth of tissue, and limb regeneration. As a result,  mechanisms for incorporating it into traditional models for cell movement, interaction, and proliferation are of great importance. A previously well-used method in order to incorporate domain growth into on-lattice reaction-diffusion models causes a build up of particles on the boundaries of the domain, which is particularly evident when diffusion is low in comparison to the rate of domain growth. Here, we present a new method which addresses this unphysical build up of particles at the boundaries, and demonstrate that it is accurate for scenarios in which the previous method fails. Further, we discuss for which parameter regimes it is feasible to continue using the original method due to diffusion dominating the domain growth mechanism.
\end{abstract}

\section{Introduction} \label{sect:Intro}


Domain growth is an inherent feature of many biological systems, from neural crest cell migration \citep{mclennan2012mmc, kulesa2010cnc} to the growth and shrinkage of tissue \citep{yates2014dcm, wolport2015pod}, and it has been investigated in the context of pattern formation in reaction-diffusion systems \citep{crampin1999rdg,crampin2000rdp}. It is therefore important that we have reliable mathematical tools in order to model such systems.


One method of modelling general reaction-diffusion systems is to compartmentalise the spatial domain into a lattice of small regions, in which particles are considered to reside. Typically, particles are permitted to jump between neighbouring compartments (although non-local jumping can also be incorporated \citep{taylor2015dab}), and particles may react with others in their current compartment (similarly, in some methods, particles may be allowed to interact with particles in neighbouring compartments \citep{isaacson2013crd}). These events, under the assumption that updates happen in continuous time, are considered Markovian and, as a result, have associated exponentially distributed waiting times. The most commonly used method to simulate such systems is the Gillespie algorithm \citep{gillespie1977ess}, however there are many others that are also used within the literature (see, for example, the next reaction method \citep{gibson2000ees}, the next subvolume method \citep{elf2004ssb, hattne2005srd}, and the sorting direct method \citep{mccollum2006sdm}).

The word `particle' in this context can refer to a multitude of biological or physical entities; a particle may be a cell when modelling biological systems such as neural crest cell migration in embryonic development, a chemical species when considering pattern formation or different classes of individual in the case of the spread of an epidemic. The discretisation of the domain does not necessarily correspond to any physical attribute of the space being modelled, but is usually used as a mathematical tool through which we are able to model stochasticity in particle positions and domain growth. As an example, consider the formation of a morphogen gradient on a growing domain \citep{smith2012ics}. In this example, the particles are the morphogen molecules, and the domain may be discretised to arbitrary accuracy, with no physical meaning necessarily being attached to the compartments. This case study will be investigated in more detail in Section \ref{sect:StretchaseStudy}.

\begin{figure}[h!!!!]
	\centering
		\includegraphics[width=0.8\textwidth, trim={0cm 4cm 0cm 3.5cm}, clip]{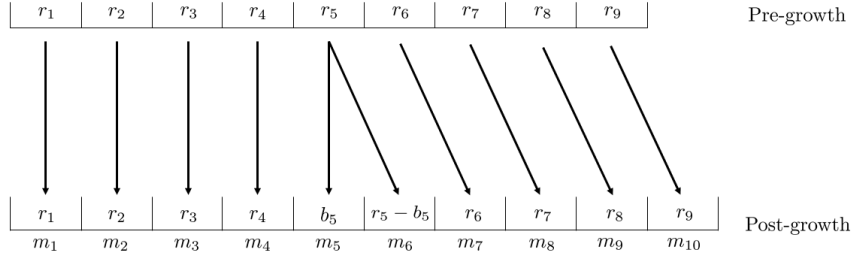}
		\caption{A schematic showing a possible splitting event when there are nine pregrowth compartments (compartments identified by the subscripts on the particle numbers). Pregrowth particle numbers are denoted $r_i$. In this case, compartment 5 is chosen to  divide, and its contents are split binomially with probability of success $1/2$ ($b_5\sim\text{Bin}(r_5,1/2)$) between the compartment in the original position (compartment 5 in this case)  and the new one that is created to its right. The compartments originally  numbered  6, 7, 8 and 9 are moved one position to the right and become  compartments 7, 8, 9 and 10, respectively. Postgrowth particle numbers are  denoted by the $m_i$'s.}
		\label{fig:Growing_Schematic}
\end{figure}

One possible option for incorporating domain growth into these on-lattice position-jump processes has been previously suggested by \citet{baker2009fmm}. Growth is achieved by choosing compartments to divide uniformly at random at a given rate. Once chosen, a compartment instantaneously doubles in length and splits down the middle to produce a new compartment (pushing all compartments to the right of the one chosen by one compartment's width). Particles that resided in the original compartment are then redistributed into the newly created compartments by placing each particle into one or the other with equal probability. This is  illustrated schematically in Figure \ref{fig:Growing_Schematic}, and a method for its implementation is detailed in Algorithm \ref{alg:Original}. A commonly implemented simulation allows particles to diffuse by jumping between neighbouring boxes while, at the same time, attempting to grow the domain uniformly \citep{baker2009fmm,yates2014dcm,yates2013ivd}. When simulating this scenario using the method of \citet{baker2009fmm} and averaging over multiple repeats, it becomes apparent that there is an issue with particles building up at the ends of the domain, an effect that can be expected when diffusion is low in comparison to the domain growth rate. When diffusion is larger, there is enough time for particles to relax to equilibrium (particles have enough time to occupy newly created sites) before the next domain growth event.

\begin{Algorithm}{The original method \citep{baker2009fmm}} \label{alg:Original}
\item Initialise time $t=0$ and set the final time $T>0$. Initialise the number of compartments $k$ and the particle numbers in each compartment $m_i,\ i\in\{1,...,k\}$. Specify the size of compartment $h$, the jump rate $d = D/h^2$ and the growth rate $\rho$.
\item Calculate the propensity functions $\alpha_i^L = dm_i$, $\alpha_i^R = dm_i$ and $\alpha_i^G = \rho$ for each compartment $i\in\{1,...,k\}$ for left jumps, right jumps and growth events respectively, with $\alpha_1^L=0=\alpha_k^R$. Calculate the sum of the propensity functions: $$\alpha_0 = \sum_{i=1}^k{\left(\alpha_i^L + \alpha_i^R + \alpha_i^G\right)}.$$ \label{alg_step:Original_Return}
\item Calculate the time until the next event by firstly drawing a uniform random variable between zero and one, $u_1\sim\text{Unif}(0,1)$, and setting $\tau = 1/\alpha_0\log\left(1/u_1\right)$. Update the time $t \leftarrow t + \tau$.
\item Determine which event, amongst all compartments, is next to occur by choosing one at random with probability proportional to the propensity function.
	\begin{enumerate}
	\item If the event corresponds to a left (resp. right) jump event from compartment $i$, remove a particle from compartment $i$ and add one to compartment $i-1$ (resp. $i+1$).
	\item If the event corresponds to a growth event in compartment $i$: Record the pregrowth particle numbers $\bs{r}\leftarrow\bs{m}$, create an extra compartment at the right end of the postgrowth domain (increasing $k$ by 1 by setting $k \leftarrow k + 1$), draw a binomial random variable with $r_i$ trials and probability of success 1/2, $b\sim\text{Bin}(r_i,1/2)$, and redistribute particles as follows (for $j\in\{1,...,k\}$): $$m_j = \left\{\begin{array}{ll}
	r_j, & \text{if }j\in\{1,...,i-1\}, \\ 
	b, & \text{if }j = i, \\ 
	r_i-b, & \text{if }j = i+1, \\ 
	r_{j-1}, & \text{if }j\in\{i+2,...,k\}.
	\end{array} 
	\right.$$
	\end{enumerate}
\item If $t < T$ then return to step \ref{alg_step:Original_Return}, else end.
\end{Algorithm}

In order to rectify this problem we have designed a novel mechanism which enables the implementation of unbiased domain growth --- one that prevents the accumulation of particles at the boundaries. This method enacts a more continuous approach to domain growth compared to \citet{baker2009fmm}, in which compartments are first stretched and then renormalised with concurrent redistribution of their particles in to neighbouring compartments. In general, we will consider growth dynamics in isolation, without considering any other mechanisms such as diffusion and reactions. This is in order to demonstrate the possible issues without confounding or hiding growth induced phenomena with other dynamics. All models will be presented in one dimension, but extensions to higher dimensions will be discussed in Section \ref{sect:Discussion}.


This paper will be set out in the following way. In Section \ref{sect:Orig} we investigate and explain the problems with the commonly employed domain growth mechanism in more detail. In Section \ref{sect:Stretch}, we present a novel, but distinct and importantly unbiased domain growth method that we use to address this issue. Finally, we present our conclusions in Section \ref{sect:Discussion}.

\section{Problems with an existing domain growth mechanism} \label{sect:Orig}

Within this section, we describe and demonstrate problems with an existing domain growth mechanism at the mesoscale \citep{baker2009fmm}, which we will refer to as the ``original method''. We then postulate why the phenomenon has not been noticed before and confirm the existence of the problem using three different techniques. 

The original method causes an accumulation of particles at the ends of the domain. This build up is caused by the inherent bias in the way in which growth is implemented: compartments are always shifted to the right when one is chosen to split. When a single compartment is chosen to split, most postgrowth compartments will either retain their pregrowth contents (if a pregrowth compartment to its right splits) or will take on the contents of the pregrowth compartment to its left (if the split is to the left of that pregrowth compartment). The notable exceptions to these rules are when postgrowth compartments retain or gain (respectively) only half of the particles (on average) from its or a neighbour's (respectively) pregrowth compartment. These events only affect two compartments per splitting event, but are nevertheless important.

The only way a postgrowth compartment retains half of its particles (on average) is if it is the compartment chosen to split. Similarly, the only way a postgrowth compartment gains half of the particles (on average) of a neighbouring compartments is if the pregrowth neighbour immediately to the left splits. All the postgrowth compartments, therefore, gain or retain half a compartments worth of particles (on average) in two different ways. The only exceptions are the first and last postgrowth compartments. The first postgrowth compartment can only retain half of its particles (on average) if it is chosen to split. It can never gain half of the particles from another compartment as there are no  pregrowth  compartments to its left that can split. Similarly, the last postgrowth compartment can never retain half of its particles since it did not exist on the pregrowth domain. Instead it can only gain half of the particles from a neighbouring compartment when the pregrowth compartment in the final position splits.

If particles are initially distributed uniformly, then an unbiased domain growth method will maintain this  uniform distribution (on average, and provided no particles enter or leave the domain). For the method of \citet{baker2009fmm} (see Figure \ref{fig:Growing_Schematic} and Algorithm \ref{alg:Original}), splitting events will redistribute the particles into the postgrowth compartments which correspond to the pregrowth compartment chosen to split and its neighbour to the right. For example, in Figure \ref{fig:Growing_Schematic}, the particle redistribution event affects postgrowth compartments 5 and 6 when pregrowth compartment 5 is chosen to split. If these splitting events affected all compartments equally then the domain growth method would be unbiased and the particle profile would remain uniform as the domain grew. However, since the two end compartments suffer (on average) only half of the particle-reducing splitting events that the non-end compartments suffer, this leads to particles accumulating at the ends of the domains.

The build up of particles at either end of the domain has previously been hidden by the smoothing effect of fast diffusion \citep{baker2009fmm}. If diffusion is large in comparison to the rate of domain growth, particles are able to diffuse away from the high concentration regions at the end of the domain, leading to a near-uniform particle profile (given a uniform initial condition and zero-flux boundary conditions). 

We will illustrate the problems with the compartment-based domain growth method  of \citet{baker2009fmm} in three different ways. For the first we undertake a stochastic simulation of the original method using the Gillespie algorithm \citep{gillespie1977ess} and average particle densities over many repeats. For the second demonstration, we calculate the numerical solution of the mean-field equations that stem from the corresponding domain growth master equation of the stochastic process in the absence of diffusion. Thirdly, we employ an analytical mathematical argument based on local redistribution of particles to assess the particle distribution in the limit of large numbers of boxes.

\begin{figure}[h!!!!]
	\begin{center}
		\subfigure[][]{
		\includegraphics[width=0.31\textwidth]{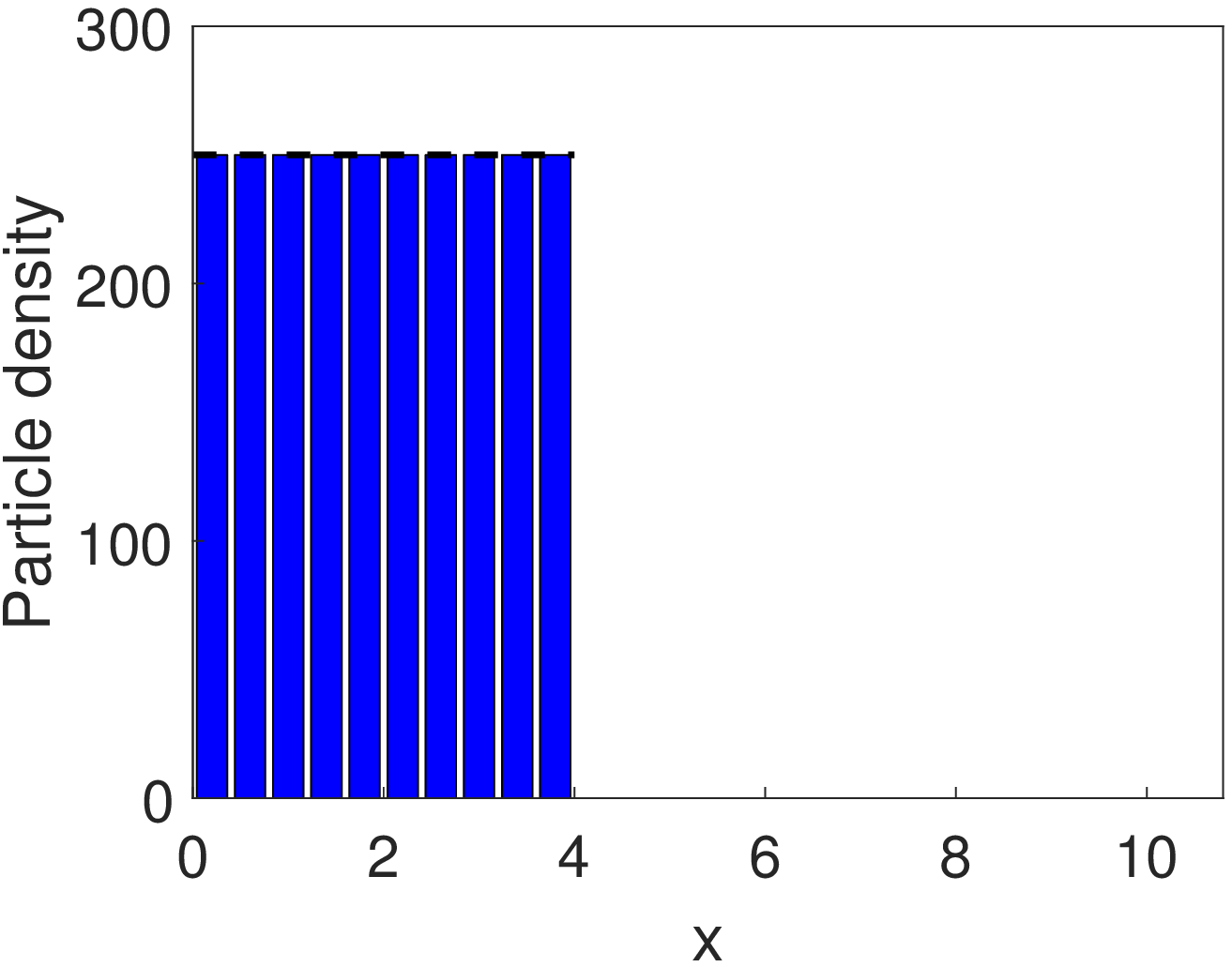}
		\label{fig:Growing_meso_toch_prob_1_time_0}}
		\subfigure[][]{
		\includegraphics[width=0.31\textwidth]{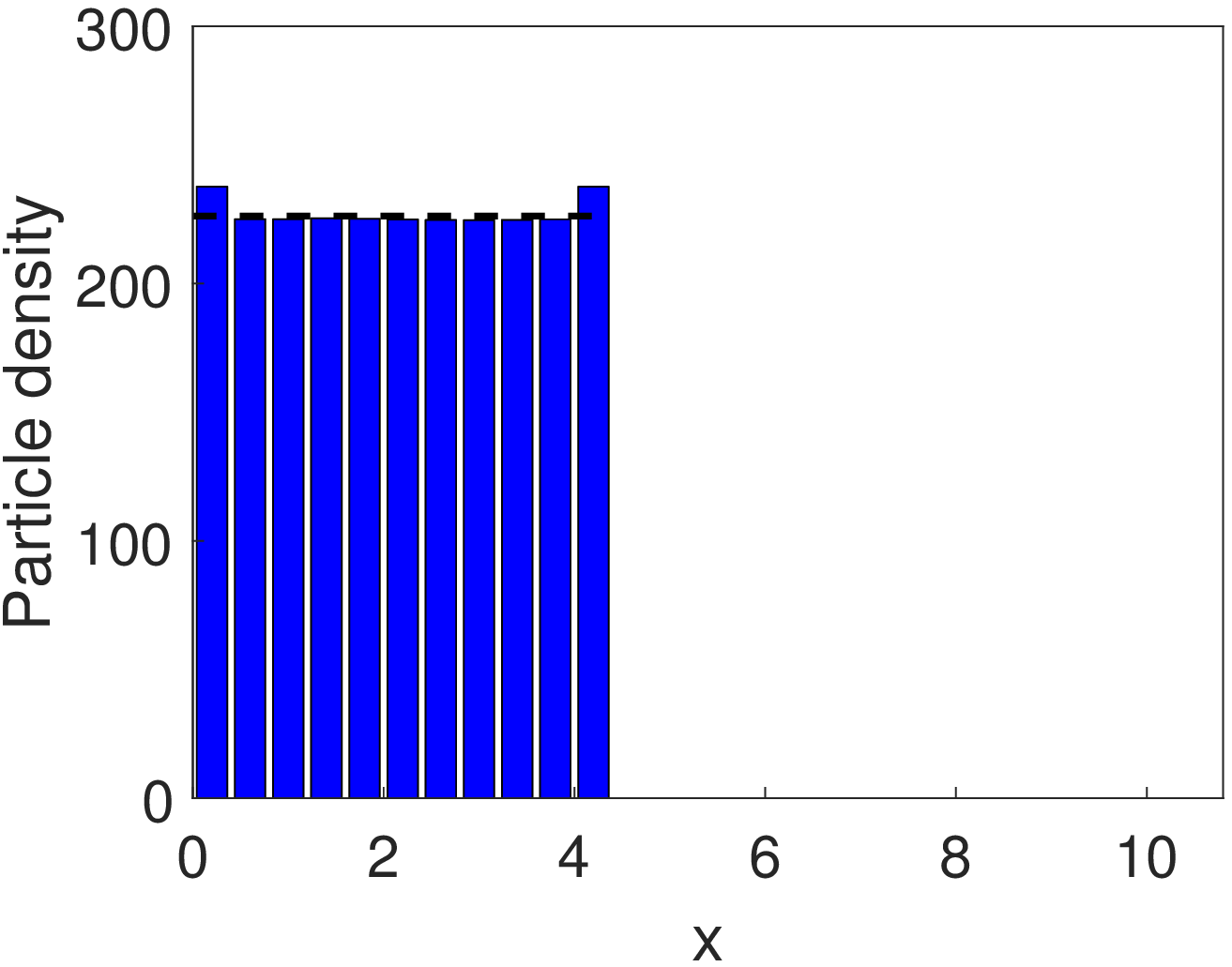}
		\label{fig:Growing_meso_toch_prob_1_time_10}}
		\subfigure[][]{
		\includegraphics[width=0.31\textwidth]{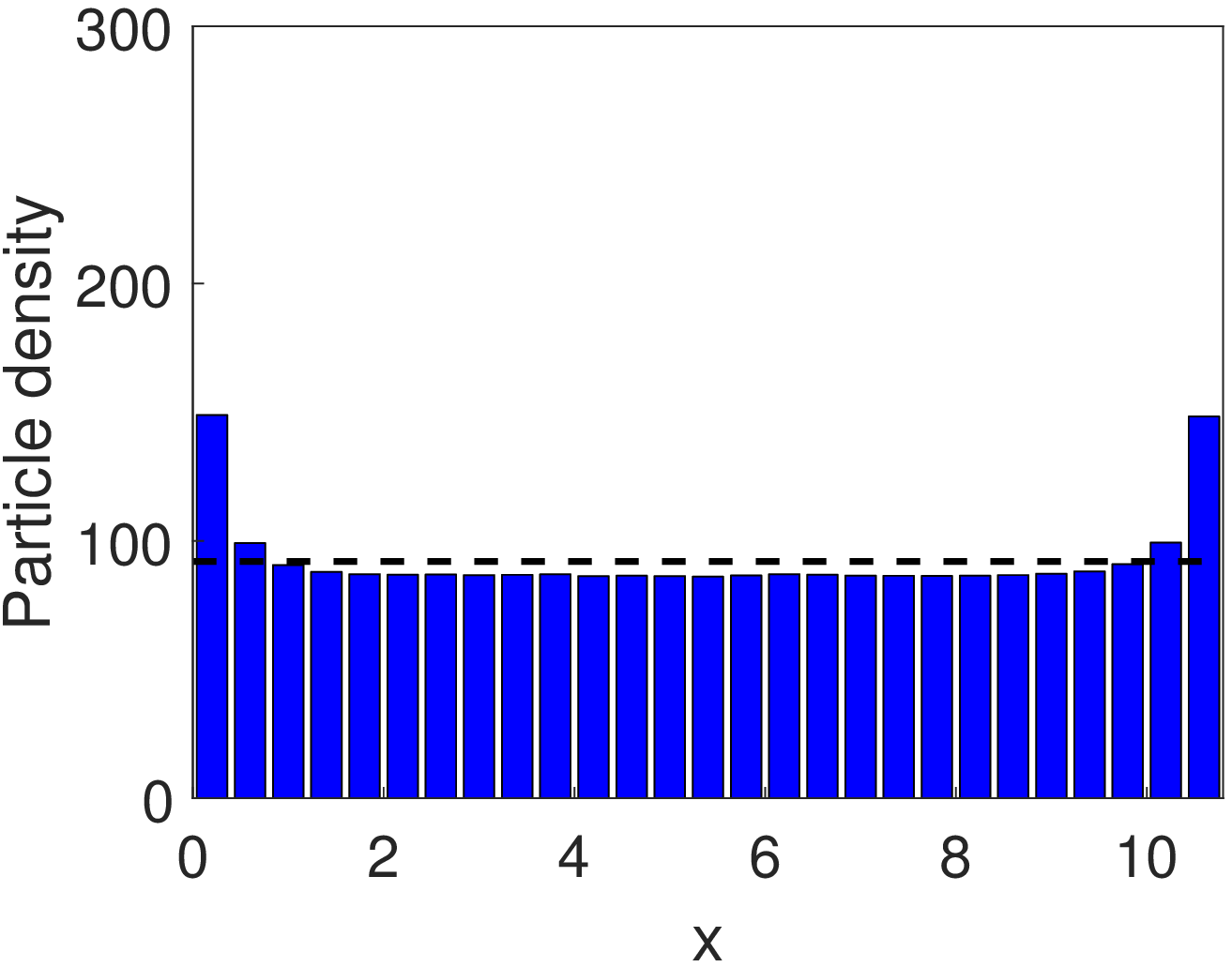}
		\label{fig:Growing_meso_toch_prob_1_time_tf}}
		\caption{Three snapshots of the stochastic simulation which demonstrate the issue with the original domain growth method \citep{baker2009fmm}. The blue bars denote particle densities in each compartment, and the black-dashed lines denote the expected behaviour of an unbiased growth method. Particle numbers are averaged over 50,000 repeats. All other parameter values are as in the text. \subref{fig:Growing_meso_toch_prob_1_time_0} The initial configuration shows 1,000 particles spread uniformly across the initial domain $[0,4]$. \subref{fig:Growing_meso_toch_prob_1_time_10} At time 10, particle build up at the domain ends is already evident. \subref{fig:Growing_meso_toch_prob_1_time_tf} Once the system has evolved to time 100, particles have accumulated significantly close to the domain ends.}
		\label{fig:Growing_meso_toch_prob_1}
	\end{center}
\end{figure}

For the stochastic simulation, we initialise a total of 1000 particles uniformly in the domain $[0,4]$ which grows exponentially in time with rate $\rho$ (we will use exponential growth as our primary test simulation, however other types of growth exhibit similar problems). We set the compartment width to be $0.4$ so that initially we have 10 compartments. The growth rate, $\rho$, is set to be 0.01 (note that all units here are arbitrary in both space and time). We choose compartment splitting times to be deterministic, meaning that we pre-calculate the times at which a compartment should split (on average) and always enact a splitting event at those times, while the Gillespie algorithm handles the stochastic events between these pre-determined times. Alternatively, growth events could be incorporated stochastically as part of the Gillespie algorithm, but for the purposes of demonstrating the problem, and for ease of visualisation, we use deterministic growth in this exposition. The compartment to divide at each predetermined splitting time is chosen uniformly at random from amongst the current compartments. Diffusion is set to be 0, so that the only effect on compartment occupancy is domain growth.
Using this domain growth configuration and averaging over 50,000 repeats, we see the clear build up of particles at both ends of the domain (see Figure \ref{fig:Growing_meso_toch_prob_1}).

To further corroborate these results and explain where this behaviour originates from, we consider the master equation for the splitting algorithm outlined above, which was first derived in \citep{baker2009fmm}. Let $p(\bs{m},k,t) = \mathbb{P}(\bs{N}(t)=\bs{m},K(t) = k)$ be the probability that the state variable (the number of particles in each compartment) at time $t$ is $\bs{m}=(m_1,...,m_k)^T$ and the number of compartments at time $t$ is $k$. This quantity evolves according to: 
\begin{equation}
\ordder{p}{t}(\bs{m},k,t) = \rho\sum_{j=1}^{k-1}{\pi(m_j,m_{j+1}|m_j+m_{j+1})p(G_j\bs{m},k-1,t)} - \rho kp(\bs{m},k,t).
\label{eqn:Growing_ME}
\end{equation} 
Here $\rho$ is the splitting rate, which is the rate at which each compartment is chosen to divide, $\pi(x,y|z)$ is a distribution describing the probability that, given there are $z$ particles in a compartment before splitting, there are $x$ and $y$ particles in the two post-split compartments (where $x+y=z$), and $G_j:\mathbb{R}^k \rightarrow \mathbb{R}^{k-1}$ is an operator that combines the contents of compartments $j$ and $j+1$ (the opposite process to splitting), so that \begin{equation*}
G_j: (m_1,...,m_j,m_{j+1},...,m_k)^T \mapsto (m_1,...,m_j+m_{j+1},...,m_k)^T.
\end{equation*} Note that the growth considered for the master equation is stochastic in both position and timing.

\begin{figure}[h!!!!]
	\begin{center}
		\subfigure[][]{
		\includegraphics[width=0.48\textwidth]{Original_method_100.eps}
	\label{fig:Growing_mean_field_Orig}}
		\subfigure[][]{
		\includegraphics[width=0.48\textwidth]{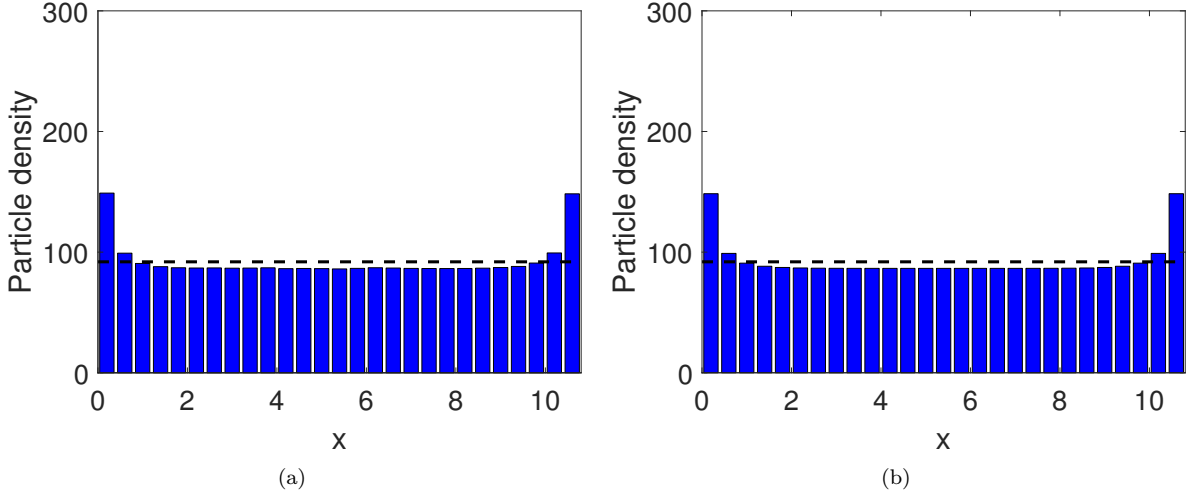}
		\label{fig:Growing_mean_field_Eqs}}
		\caption{A comparison between \subref{fig:Growing_mean_field_Orig} the particle density found by using Algorithm \ref{alg:Original}, averaged over 50,000 repeats (reproduced from Figure \ref{fig:Growing_meso_toch_prob_1_time_tf}), and \subref{fig:Growing_mean_field_Eqs} the corresponding particle density found by solving mean-field equations \eqref{eqn:Growing_MF}. The mean-field equations are initialised in the same way as the stochastic algorithm (see Figure \ref{fig:Growing_meso_toch_prob_1_time_0} and related caption).} 
		\label{fig:Growing_mean_field}
	\end{center}
\end{figure}

We can calculate the mean-field equations for the evolution of the mean number of particles in each compartment, under the splitting probability corresponding to a symmetric binomial distribution $$\pi(x,y|z) = {z\choose x}\left(\frac{1}{2}\right)^z,$$ by multiplying both sides of equation \eqref{eqn:Growing_ME} by $m_i$ for each index $i\in\{1,...,k\}$, and summing over the entire postgrowth state variable. Define the quantity 
\begin{equation*}
M_i^k(t) = \sum_{\bs{m}\in \mathbb{N}_0^k}{m_ip(\bs{m},k,t)},
\end{equation*}
which is the mean number of particles in compartment $i$ at time $t$, given that there are $k$ compartments in the system overall. These corresponding mean-field equations are given by: \begin{equation}
\ordder{M_i^k}{t}(t) = \rho\left[\left(i-\frac{3}{2}\right)M_{i-1}^{k-1}(t) + \left(k-\frac{1}{2}-i\right)M_i^{k-1}(t) - kM_i^k(t)\right],
\label{eqn:Growing_MF}
\end{equation} which holds for $k\in\mathbb{N},\ j\in\{1,...,k\}$ and $t>0$ \citep{baker2009fmm}, noting that $M_k^{k-1}\equiv 0$. The full derivation is omitted here, however we direct the interested reader to \citet{baker2009fmm}, or Appendix \ref{sect:Appendix_Stretch_Mean} for a similar calculation. We plot the solutions to equation \eqref{eqn:Growing_MF} in order to illustrate the average behaviour of the system under the biased domain growth algorithm, in Figure \ref{fig:Growing_mean_field}. We show that in this case, we exhibit the same build up of particles at the boundaries that is evident in the stochastic simulation, shown in Figure \ref{fig:Growing_meso_toch_prob_1}.

Finally, we gain a more quantitative insight into the bias engendered by the domain growth method of \citet{baker2009fmm} by using an analytical approach. In particular, we derive coefficients which describe the densities of  particles in each compartment. We let $u_i^k$ denote the normalised density of particles in compartment $i$ when there are $k$ compartments in total (a random variable), $M_i^k = \mathbb{E}[u_i^k]$, and $I^k$ is the index of the compartment which splits when there are $k$ compartments in total (chosen uniformly at random for each event). Finally, we define $q$ to be a sample from some symmetric distribution with mean $1/2$. This will denote the random proportion of density that is placed in the left-hand postgrowth compartment when there are $k$ compartments in total, and is independent of the particle density and the compartment chosen to split.
We begin with all of the density in the first (and only) compartment, so that $u_1^1 = 1$. We will set up recursion relationships between the $u_i^k$'s for different $i$ and $k$ in order to approximate $u_i^k$ for large $k$. Specifically, we express $u_i^{k+1}$ in terms of $u_j^k$ for $j\in\{1,...,k\}$: 
\begin{equation}
u_i^{k+1} = \underbrace{\vphantom{q}u_i^k\mathbbm{1}_{[I^k > 
i]}}_{\substack{\text{Compartment} \\ \text{to right of } \\ i\text{ splits}}} + \underbrace{qu_i^k\mathbbm{1}_{[I^k=i]}}_{\substack{\text{Compartment } \\ i\text{ splits}}} + \underbrace{(1-q)u_{i-1}^k\mathbbm{1}_{[I^k=i-1]}}_{\substack{\text{Compartment } \\ i-1\text{ splits}}} + \underbrace{\vphantom{q}u_{i-1}^k\mathbbm{1}_{[I^k < i-1]}}_{\substack{\text{Compartment} \\ \text{to left of } \\ i-1\text{ splits}}}, \qquad i\in\{2,...,k\},
\label{eqn:Asymp_gen}
\end{equation}
with similar expressions for $i=1$ and $k+1$. Here $\mathbbm{1}_{\text{[condition]}}$ is the indicator function, which is unity when the 
subscripted condition is satisfied and zero otherwise.

Considering the first compartment, relationship \eqref{eqn:Asymp_gen} stipulates that \begin{equation*}
u_1^{k+1} = u_1^k\mathbbm{1}_{[I^k>1]} + qu_1^k\mathbbm{1}_{[I^k=1]}.
\end{equation*} 
Taking expectations of this expression, noting that the choice of compartment, the density in each compartment and the proportion of the density placed in the left pregrowth compartment are independent (so that $u_i^k$, $I^k$ and $q$ are independent for every $i\in\{1,...,k\}$) we find that: \begin{align*}
M_1^{k+1} &= M_1^k\mathbb{P}(I^k>1) + \frac{1}{2}M_1^k\mathbb{P}(I^k=1),\\
&=M_1^k\left(\frac{k-1}{k}+\frac{1/2}{k}\right),\\
&=M_1^k\left(1-\frac{1/2}{k}\right).
\end{align*} Applying this recursive relation $k$ times: \begin{equation}
M_1^{k+1} = M_1^1\prod_{n=1}^{k}{\left(1-\frac{1/2}{n}\right)}=\prod_{n=1}^{k}{\left(1-\frac{1/2}{n}\right)}, \label{eqn:asymp_n1}
\end{equation} since $M_1^1 = \mathbb{E}[u_1^1] = 1$. In order to simplify this expression, we can use the following relationship, which is derived in Appendix \ref{sect:appendix_1}:\begin{equation}
\prod_{n=a}^b{\left(1+\frac{c}{n}\right)} \approx \frac{\Gamma(a)}{\Gamma(a+c)e^c}b^c,
\label{eqn:asymp_approx}
\end{equation} where $a,b\in\mathbb{N}$ such that $a<b$ and $b$ is large, and $c\in\mathbb{R}$. Applying approximation \eqref{eqn:asymp_approx} to equation \eqref{eqn:asymp_n1} for large $k$ gives: \begin{equation}
M_1^{k+1} = \prod_{n=1}^{k}{\left(1-\frac{1/2}{n}\right)} \approx  \frac{\Gamma(1)}{\Gamma(1/2)e^{-1/2}}k^{-1/2} = \frac{1}{\sqrt{\pi}e^{-1/2}}k^{-1/2}. \label{eqn:asymp_n1_sol}
\end{equation}
Now consider the second compartment. As with the first, we write the recursion relation \eqref{eqn:Asymp_gen}: \begin{equation*}
u_2^{k+1} = u_2^k\mathbbm{1}_{[I^k>2]} + \frac{1}{2}u_2^k\mathbbm{1}_{[I^k=2]} + \frac{1}{2}u_1^k\mathbbm{1}_{[I^k=1]}.
\end{equation*} Once again, we take expectations and simplify by applying relation \eqref{eqn:Asymp_gen} recursively (as in equation \eqref{eqn:asymp_n1}):\begin{align}
M_2^{k+1} &= M_2^k\mathbb{P}(I^k>2) + \frac{1}{2}M_2^k\mathbb{P}(I^k=2) + \frac{1}{2}M_1^k\mathbb{P}(I^k=1),\notag\\
&\vdots\notag\\
&= M_2^2\underbrace{\prod_{n=2}^k{\left(1-\frac{3/2}{n}\right)} \vphantom{\left(\prod_{j=k+1}^n{\left(1-\frac{3/2}{j}\right)}\right)}}_{\text{Term 1}} + \underbrace{\sum_{n=2}^{k-1}{\left(\prod_{j=n+1}^k{\left(1-\frac{3/2}{j}\right)}\right)M_1^n\left(\frac{1/2}{n}\right)}}_{\text{Term 2}} + \underbrace{M_1^k\left(\frac{1/2}{k}\right)\vphantom{\left(\prod_{j=k+1}^n{\left(1-\frac{3/2}{j}\right)}\right)}}_{\text{Term 3}}.
\label{eqn:Asymp_Box_2}
\end{align} 
We consider each of the three terms in equation \eqref{eqn:Asymp_Box_2} sequentially. Using equation \eqref{eqn:asymp_approx}, we can approximate term 1, for large $k$, as \begin{equation}
\prod_{n=2}^k{\left(1-\frac{3/2}{n}\right)} \approx 
\frac{\Gamma(2)}{\Gamma(1/2)e^{-3/2}}k^{-3/2}.
\label{eqn:terms1_simplified}
\end{equation}
It can be shown (see Appendix \ref{sect:appendix_2}) that for large $k$, term 2 can be approximated by 
\begin{equation}
\sum_{n=2}^{k-1}{\left(\prod_{j=n+1}^k{\left(1-\frac{3/2}{j}\right)}
\right)M_1^n\left(\frac{1/2}{n}\right)} \approx \frac{1}{2}\frac{1}{\sqrt{\pi}e^{-1/2}}k^{-1/2} - \frac{1}{\sqrt{\pi}e^{-1/2}}k^{-3/2}.
\label{eqn:terms2_simplified}
\end{equation}
To simplify term 3 we make use of result \eqref{eqn:asymp_n1_sol}: 
\begin{align}
M_1^k\left(\frac{1/2}{k}\right) \approx \frac{1}{2}\frac{1}{\sqrt{\pi}e^{-1/2}}k^{-3/2}
\label{eqn:terms3_simplified}
\end{align}
Substituting the resultant expressions \eqref{eqn:terms1_simplified}, \eqref{eqn:terms2_simplified} and \eqref{eqn:terms3_simplified} into \eqref{eqn:Asymp_Box_2} gives, for large $k$:
\begin{align*}
M_2^{k+1} \approx M_2^2\frac{\Gamma(2)}{\Gamma(1/2)e^{-3/2}}k^{-3/2} + \frac{1}{2}\frac{1}{\sqrt{\pi}e^{-1/2}}k^{-1/2} - \frac{1}{\sqrt{\pi}e^{-1/2}}k^{-3/2} + \frac{1}{2}\frac{1}{\sqrt{\pi}e^{-1/2}}k^{-3/2},
\end{align*}
For large $k$, the $O(k^{-1/2})$ terms dominate leaving us with the following approximation:
\begin{equation}
M_2^{k+1} \approx \frac{1}{2}\frac{1}{\sqrt{\pi}e^{-1/2}}k^{-1/2}.
\end{equation}

Following the same procedure, we can find the approximate expressions for the asymptotic particle density in each of the compartments. In particular, it can be shown that 
\begin{equation}
M_i^k \approx c_ik^{-1/2},
\label{eqn:Mathematical_Final}
\end{equation} 
where 
\begin{align*}
c_1 &= \frac{1}{\sqrt{\pi}e^{-1/2}},\\
c_i &= c_1\prod_{j=2}^{i}{\frac{2j-3}{2j-2}}, \quad i \in \{2,3,...\}.
\end{align*}

%

\begin{figure}[h!!!!]
	\begin{center}
		\subfigure[][]{
		\includegraphics[width=0.48\textwidth]{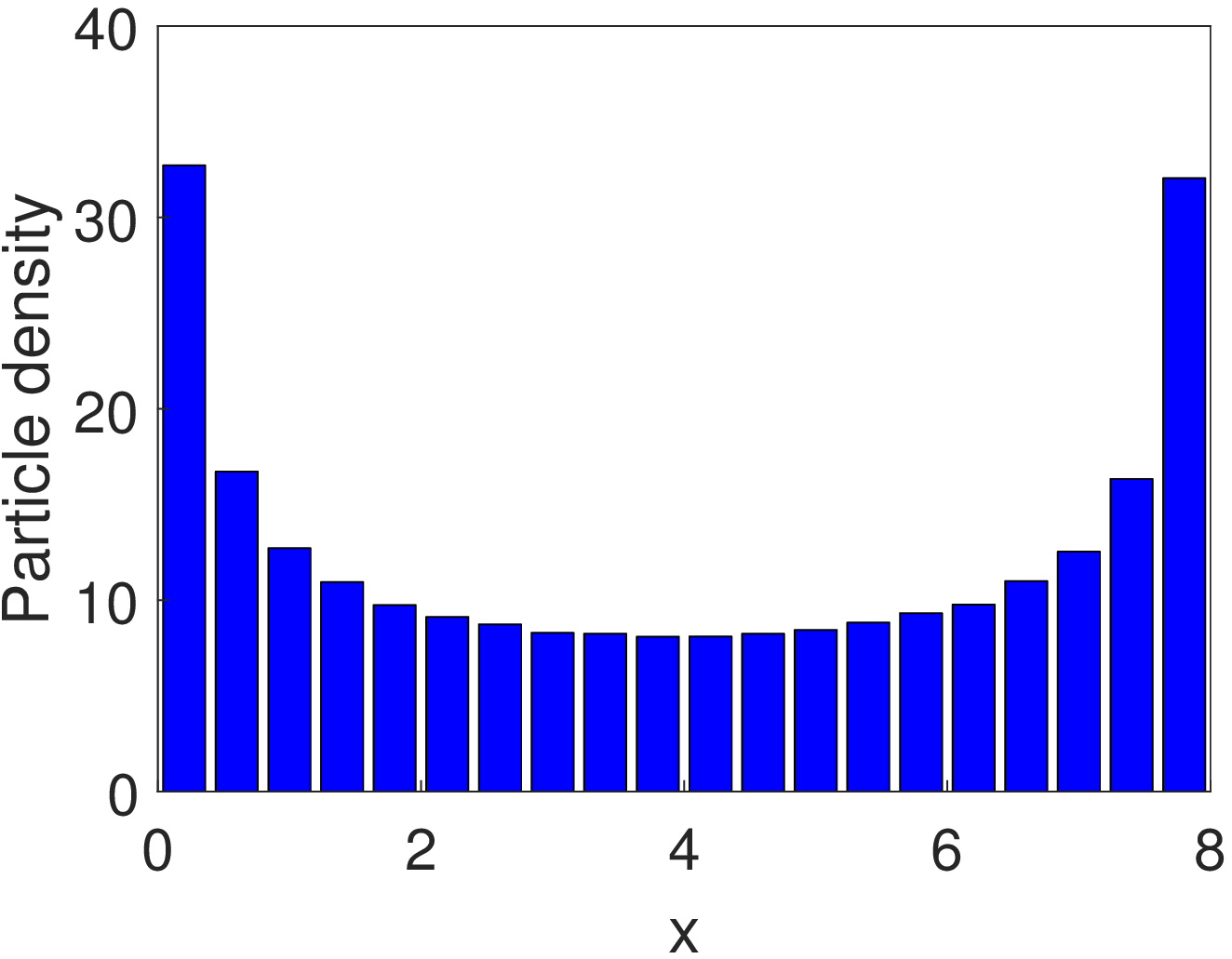}
		\label{fig:Mathematical_Plot_Stoch}}
		\subfigure[][]{
		\includegraphics[width=0.48\textwidth]{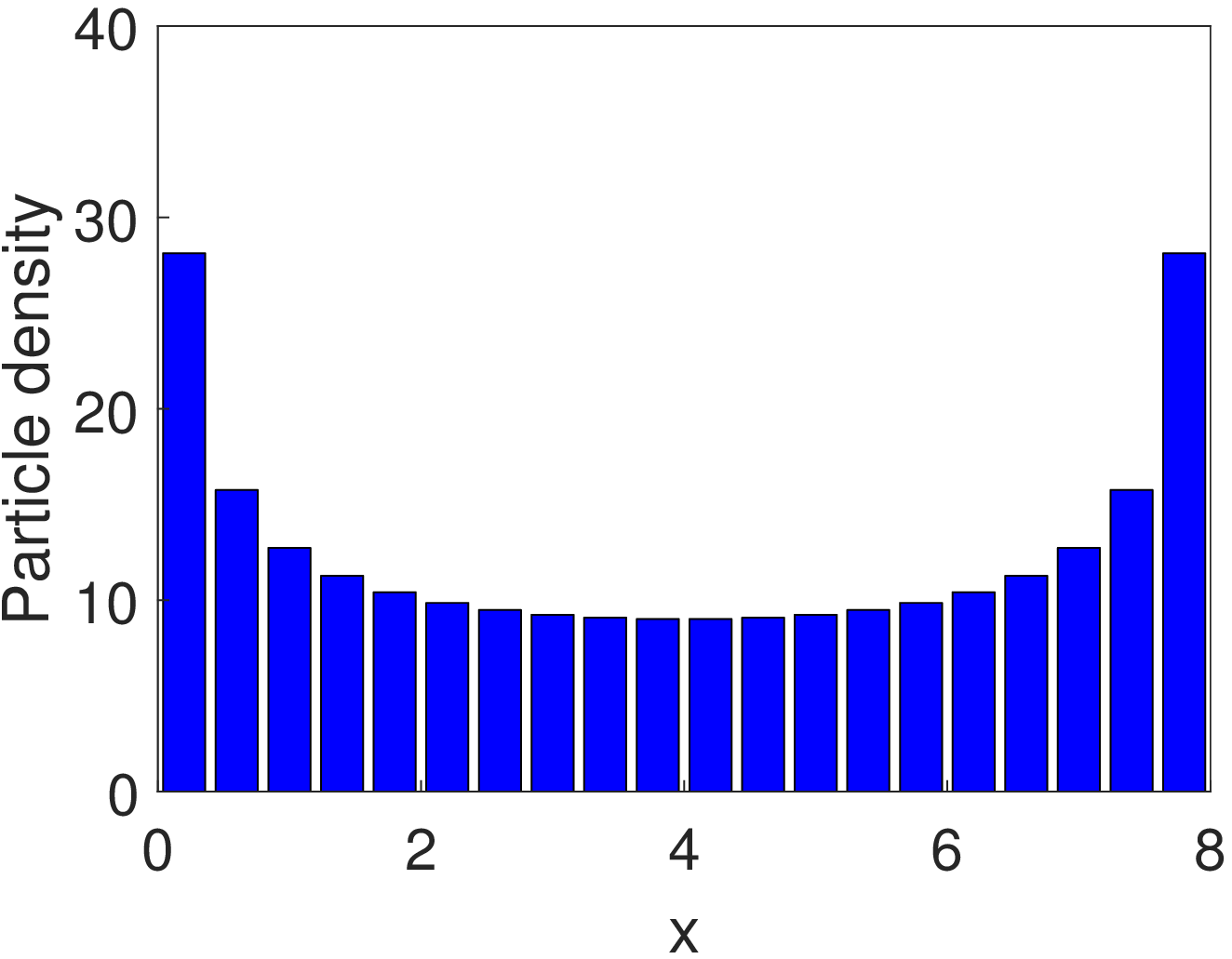}
		\label{fig:Mathematical_Plot_Maths}}
		\caption{A comparison between \subref{fig:Mathematical_Plot_Stoch} a stochastic simulation averaged over 10,000 repeats and initialised with a single compartment and \subref{fig:Mathematical_Plot_Maths} the result of the mathematical argument, where the displayed compartment $i$ is given by the average of $M_i^{20}$ and $M_{21-i}^{20}$, calculated using equation \eqref{eqn:Mathematical_Final}. Both plots have the same total area.}
		\label{fig:Mathematical_Plot}
	\end{center}
\end{figure}

To assess the accuracy of this approximation, we undertake a stochastic simulation initialised with a single compartment containing 100 particles. At time $t=100$, under our time-deterministic splitting mechanism each simulation finishes with 20 compartments. We then compare the particle numbers in each compartment, averaged over 10,000 repeats, to equation \eqref{eqn:Mathematical_Final}. Since the simulation domain has only finitely many compartments but our mathematical analysis considers an infinite number of compartments, we average $M_i^{20}$ and $M_{21-i}^{20}$ when we plot compartment $i$ under the assumption that densities are initialised, and subsequently remain, symmetric. Finally, we scale each of the plots so that they have the same number of particles. Although a quantitative agreement is not expected, since our results hold strictly only on an infinite domain, the results in Figure \ref{fig:Mathematical_Plot} demonstrate that our mathematical analysis matches the simulation results qualitatively.

The original domain growth method proposed by \citet{baker2009fmm} has been used in many compartment-based studies of domain growth \citep{woolley2011srd,thompson2012mcm,yates2012gfm,yates2014dcm}. Despite not having previously been evident, we have been able to demonstrate in three distinct ways, that this domain growth method is biased. The consequence of this bias is that particles tends to accumulate at the extreme ends of the domain. In the next section, we introduce the stretching method, which prevents this build up of particles and gives genuinely uniform, unbiased domain growth.

\section{Stretching method} \label{sect:Stretch}

We now introduce the stretching method. This differs from the original method because it is a global method as opposed to a local one. That is, instead of choosing a single compartment to instantaneously grow to twice its length and split, we stretch all compartments by a small amount and redistribute particles amongst all compartments (for a brief discussion of local growth mechanisms, please see Appendix \ref{sect:Appendix_Local}). We will firstly explain the method, before demonstrating its effectiveness. We do this by showing that it can correctly maintain a uniform particle profile on a uniformly growing domain. We then look at a case study, the formation of a morphogen gradient on an exponentially growing domain, in order to directly compare the original and stretching methods using an example with its roots in biology (see Section \ref{sect:StretchaseStudy}). Finally, in Section \ref{sect:Stretch_comparison}, we investigate the parameter regimes in which the spatial inhomogeneities in the original method are negligible.

\begin{figure}[h!!!!!!!!!]
	\centering
		\includegraphics[width=0.9\textwidth]{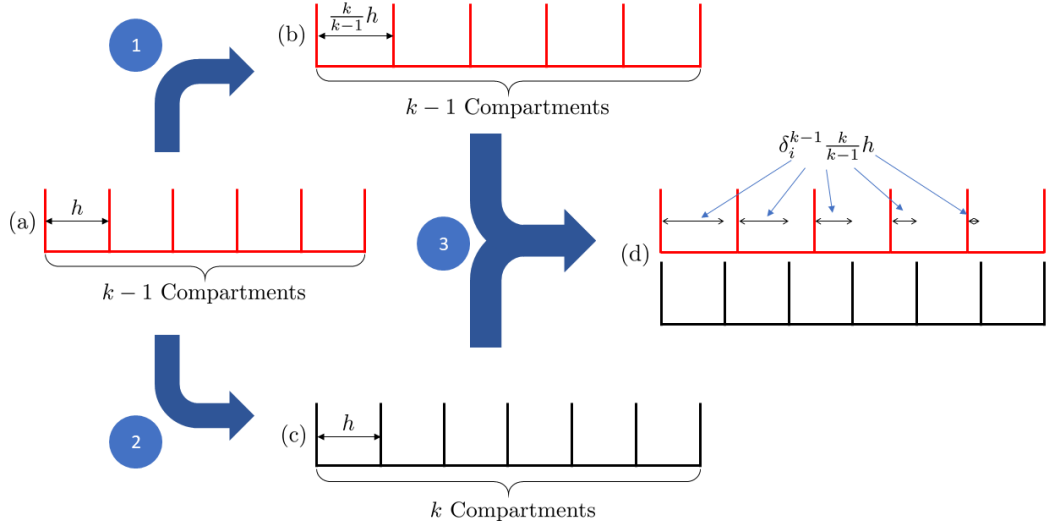}
		\caption{The process of domain growth for the stretching method. We start with $k-1$ compartments (a), and in step 1, stretch each of these compartments to be $k/(k-1)$ times their original length (b). This has the effect of increasing the domain length by $h$, a compartment's width. In step 2, we add a new compartment to the original $k-1$ compartments of size $h$ (c), which again yields a domain that is a compartment's width bigger. In step 3, we redistribute the particles in the stretched $k-1$ compartments (c) into the $k$ compartments of size $h$ (b) by calculating the $\delta_i^{k-1}$ values, which tell us how much overlap there is between the two meshes (d). These $\delta_i^{k-1}$ values are treated as the probability for each particle in a stretched compartment to move to the renormalised compartment which overlaps its left-hand boundary. If a particle does not move to the renormalised compartment which overlaps with its left hand boundary then it moves to the renormalised compartment which overlaps with its right hand boundary.}
\label{fig:Growing_stretch}
\end{figure}

We begin by describing the method. Assume the number of compartments before a growth event is $k-1$ for some $k>2$, and define each compartment to be of width $h$ (see Figure \ref{fig:Growing_stretch}(a)). We will define the state variable before growth to be $\bs{r}\in\mathbb{N}_0^{k-1}$ and after growth to be $\bs{m}\in\mathbb{N}_0^k$ in order to be consistent with the original method. The method proceeds as follows: 
\begin{enumerate}
\item When a growth event is chosen to occur, we stretch the domain to be of size $kh$ rather than $(k-1)h$ (see Figure \ref{fig:Growing_stretch}(b)). We do this uniformly across the entirety of the domain, so that each compartment on the stretched domain is now of width $kh/(k-1)$.
\item In the second step we add a compartment to the right-hand end of the pre-stretched domain (see Figure \ref{fig:Growing_stretch}(c)). It is on this postgrowth domain that we define the state variable $\bs{m}$. Note that we now have two domains, each with a different number of compartments, but both of the same length.
\item For the third step, we compare the two meshes. Note that for every stretched compartment, exactly two of the postgrowth compartments intersect it (see Figure \ref{fig:Growing_stretch}(d)). Assuming particles are uniformly distributed across each stretched compartment, we can calculate the proportion of particles, $\delta_i^{k-1}$, that should be placed in the left overlapping compartment. If we denote the right-hand end of compartment $i$ in the renormalised domain as $x_i=ih$ and use $\tilde{x}_i=\left[ki/(k-1)\right]h$ the quantity for the stretched domain, then: \begin{align}
\delta_i^{k-1}\frac{k}{k-1}h &= x_i - \tilde{x}_{i-1}\notag\\
&=ih-\frac{k}{k-1}ih + \frac{k}{k-1}h\notag\\
&= \frac{k-i}{k-1}h.\notag\\
\Longrightarrow \delta_i^{k-1} & = \frac{k-i}{k}.\label{eqn:Overlap_Proportion}
\end{align}
\item Finally we calculate the new state $\bs{m}$ by drawing $k-1$ binomial random variables $b_i \sim \text{Bin}(r_i,\delta_i^{k-1})$ and calculating $\bs{m}$ as: \begin{equation*}
\bs{m} = \sum_{i=1}^{k-1}{b_i\bs{e}_i + (r_i-b_i)\bs{e}_{i+1}},
\end{equation*} where the $\bs{e}_i$ are the standard $k$-dimensional basis vectors.
\end{enumerate}

\begin{Algorithm}{The stretching method} \label{alg:Stretch}
\item Initialise time $t=0$ and set the final time $T>0$. Initialise the number of compartments, $k$, and the particle numbers in each compartment $m_i,\ i\in\{1,...,k\}$. Specify the size of compartment $h$, the jump rate $d = D/h^2$ and the growth rate $\rho$.
\item Calculate the propensity functions $\alpha_i^L = dm_i$, $\alpha_i^R = dm_i$ for left and right jumps from each compartment for $i=1,...,k$ with $\alpha_1^L=0=\alpha_k^R$, and set the propensity function for a growth event to be $\alpha^G = \rho k$. Calculate the sum of the propensity functions: $$\alpha_0 = \sum_{i=1}^k{\left(\alpha_i^L + \alpha_i^R\right)} + \alpha^G.$$ \label{alg_step:Stretch_Return}
\item Calculate the time until the next event by firstly drawing $u_1\sim\text{Unif}(0,1)$ and setting $\tau = 1/\alpha_0\log\left(1/u_1\right)$. Set the time to be the next event that occurs $t \leftarrow t + \tau$.
\item Determine which event is next to occur by choosing one at random with probability proportional to the propensity function.
	\begin{enumerate}
	\item If the event corresponds to a left (resp. right) jump event from box $i$, remove a particle from compartment $i$ and add one to compartment $i-1$ (resp. $i+1$).
	\item If the event corresponds to a growth event: firstly, define the pregrowth state to be $\bs{r}\leftarrow\bs{m}$, calculate the overlap proportions $\delta_i^k = (k+1-i)/(k+1)$, and use these in order to draw $k$ binomial random variables $b_i\sim\text{Bin}(m_i,\delta_i^k)$. Create an extra compartment at the right end of the postgrowth domain (increasing $k$ by 1 by setting $k \leftarrow k + 1$). Then set, for $j\in\{1,...,k\}$: $$m_j = \left\{\begin{array}{ll}
	b_1, & \text{if }j = 1, \\ 
	b_j + (r_{j-1}-b_{j-1}), & \text{if }j\in\{2,...,k-1\}, \\ 
	r_{k-1}-b_{k-1}, & \text{if }j = k.
	\end{array} 
	\right.$$ 
	\end{enumerate}
\item If $t < T$ then return to step \ref{alg_step:Stretch_Return}, else end.
\end{Algorithm}

\begin{figure}[h!!!!!!!]
	\begin{center}
		\subfigure[][]{
		\includegraphics[width=0.31\textwidth]{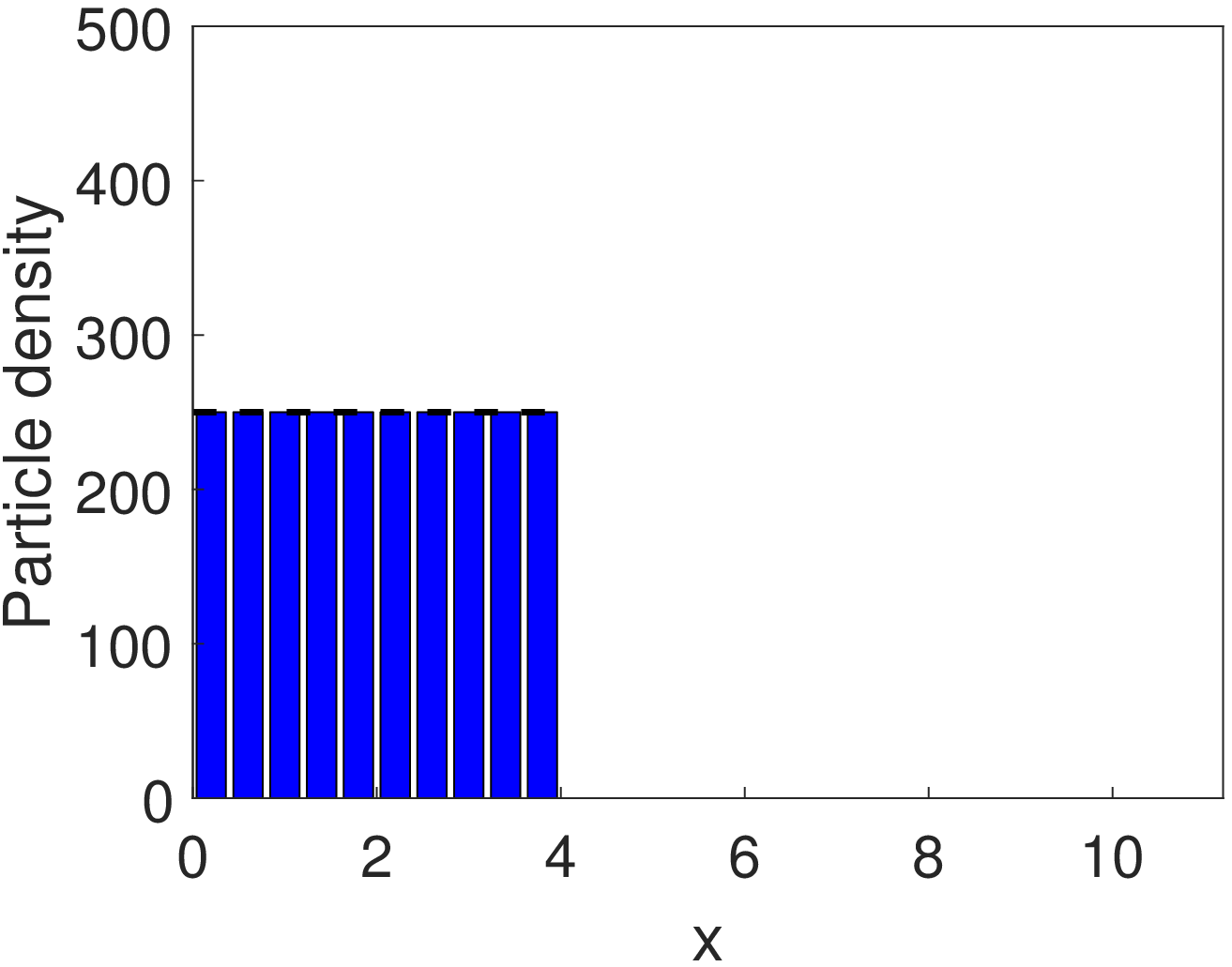}
		\label{fig:Growing_meso_stretch_0}}
		\subfigure[][]{
		\includegraphics[width=0.31\textwidth]{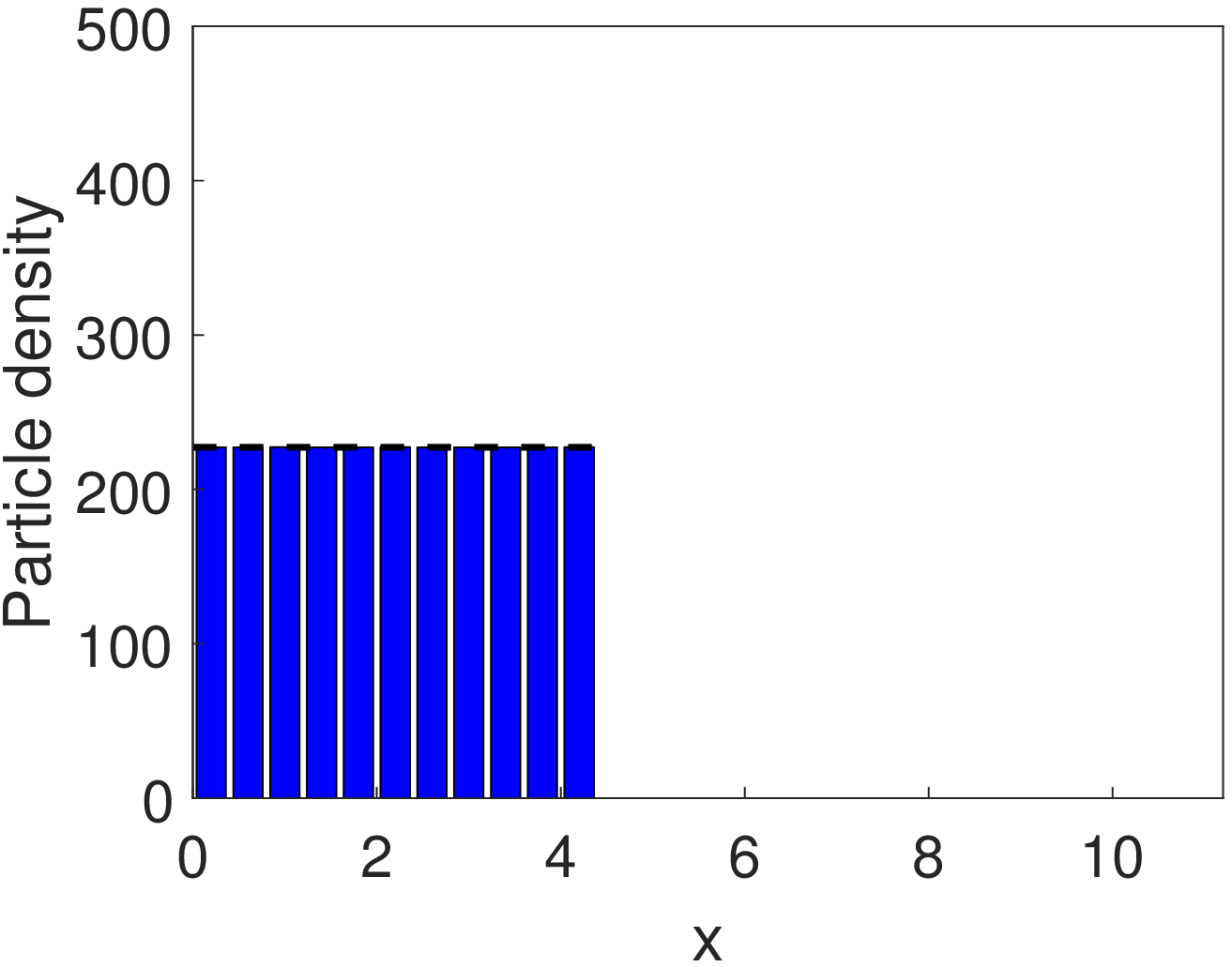}
		\label{fig:Growing_meso_stretch_mid}}
		\subfigure[][]{
		\includegraphics[width=0.31\textwidth]{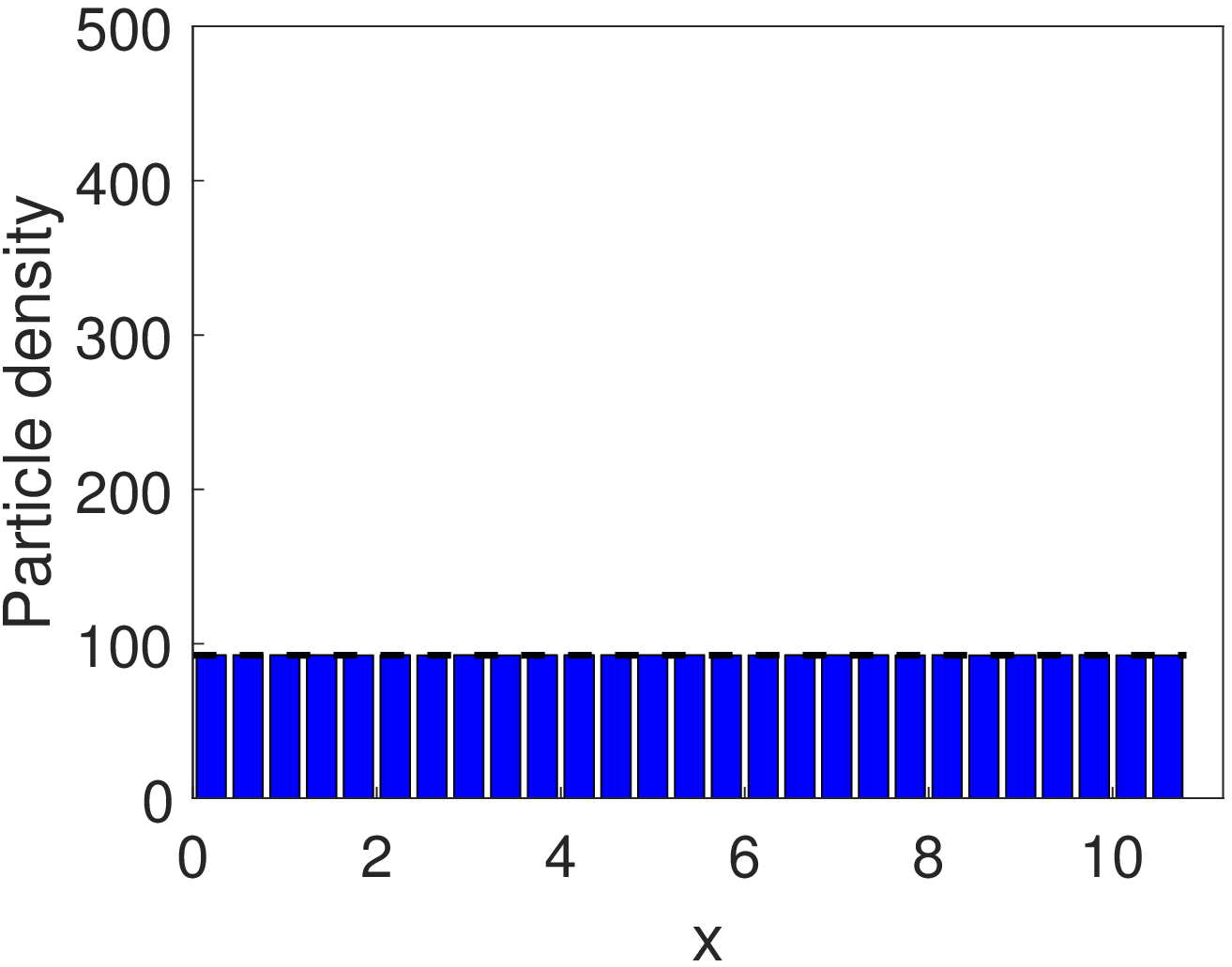}
		\label{fig:Growing_meso_stretch_tf}}
		\caption{Three snapshots of the particle density on a domain growing according to the stretching method \subref{fig:Growing_meso_stretch_0} initially, at time 0, \subref{fig:Growing_meso_stretch_mid} at time 10 and \subref{fig:Growing_meso_stretch_tf} at the final time 100. All descriptions and parameter values are the same as in Figure \ref{fig:Growing_meso_toch_prob_1}.}
		\label{fig:Growing_meso_stretch}
	\end{center}
\end{figure}

We assess the stretching method by initialising a uniform profile and test to see whether uniformity is maintained under the stretching domain growth method. In Figure \ref{fig:Growing_meso_stretch}, we can see that the stretching method performs very well in comparison to the original method.

\subsection{Case study: Morphogen gradient} \label{sect:StretchaseStudy}

For our case study, we apply the original and stretching methods to the formation of a morphogen gradient on an exponentially growing domain \citep{wolpert1969pis, painter1999sfj}. We begin with an initial domain of length $L_0$, which grows with rate $\rho$, and on which particles with density $u(x,t)$ move and interact. Particles diffuse with diffusion coefficient $D$, and they decay uniformly at a rate $\mu$. There is also an influx of particles at the left-hand boundary, at rate $D\lambda$. We will only allow particles to jump between adjacent compartments and to interact within their own compartments. In order to compare the results of the stochastic simulation, we compute the solution to the associated mean-field PDE, which is given by: \begin{align}
\text{PDE} &: \partder{u}{t} = D\secpartder{u}{x} - \partder{(\rho xu)}{x} - \mu u, && x\in(0,L_0\exp\left(\rho t\right)),\ t>0,\notag\\
\text{BC} &: \left.\partder{u}{x}\right|_{x=0}=-\lambda,\ \left.\partder{u}{x}\right|_{x=L_0\exp\left(\rho t\right)}=0, && t > 0,\label{eqn:Morph_PDE}\\
\text{IC} &: u(x,0) = 0, && x\in [0,L_0].\notag
\end{align} The PDE comprises three terms on the right-hand side --- the first is the diffusive term, the second is dilution due to domain growth and the third is degradation of particles over the spatial domain \citep{crampin1999rdg}. The first boundary condition is the in-flux of particles at the left of the domain, while the second boundary condition is a reflective boundary. For our on-lattice simulations, diffusion is implemented by allowing particles to jump between neighbouring compartments with rate $d = D/h^2$. Particle degradation is achieved by allowing each particle in a compartment to be removed with rate $\mu$, while production is included through a production reaction in the compartment adjacent to the left-hand boundary, as is described by \citet{taylor2015dab}.

\begin{figure}[h!!!!]
	\begin{center}
		\subfigure[][]{
		\includegraphics[width=0.31\textwidth]{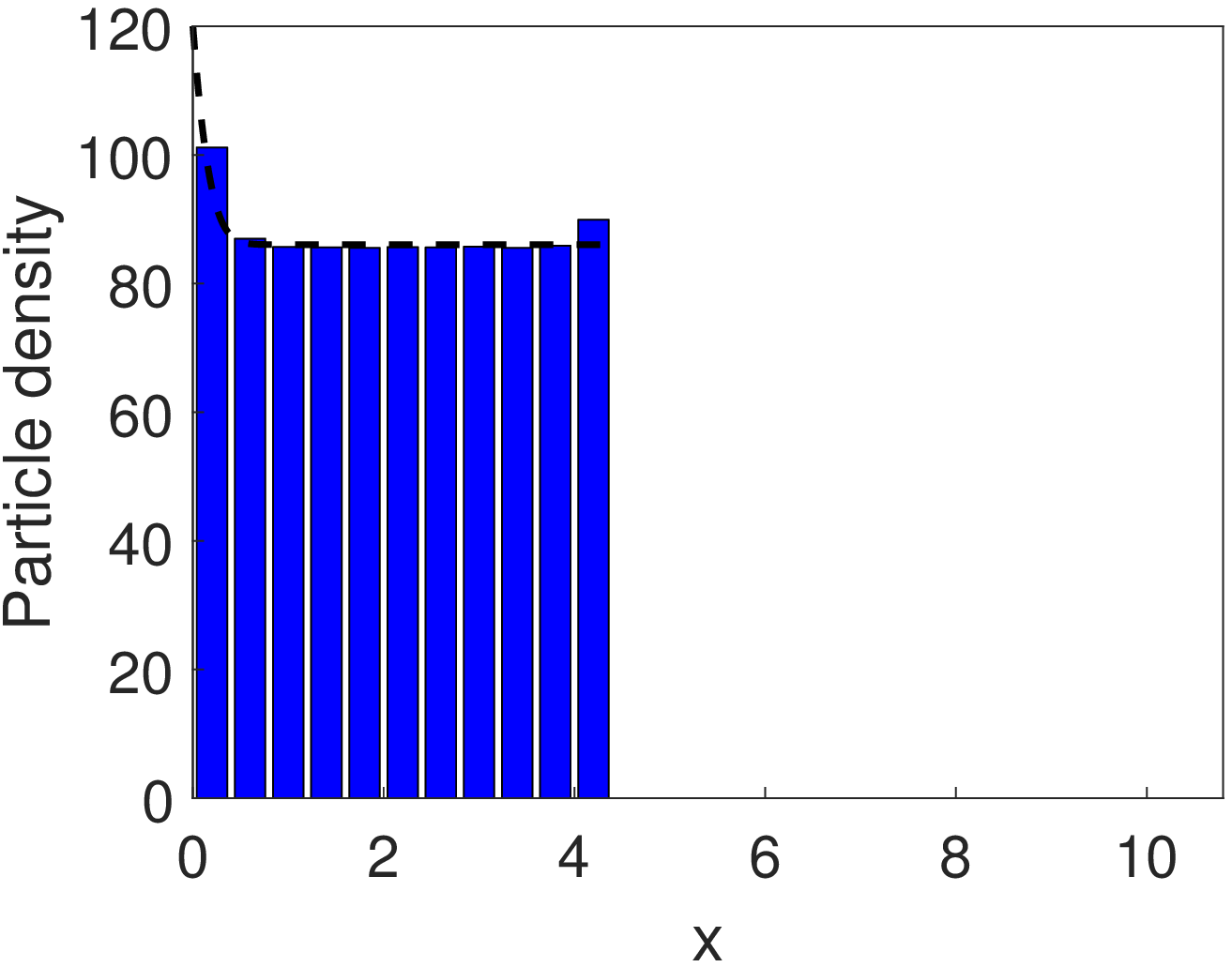}
		\label{fig:Morphogen_orig_mid}}
		\subfigure[][]{
		\includegraphics[width=0.31\textwidth]{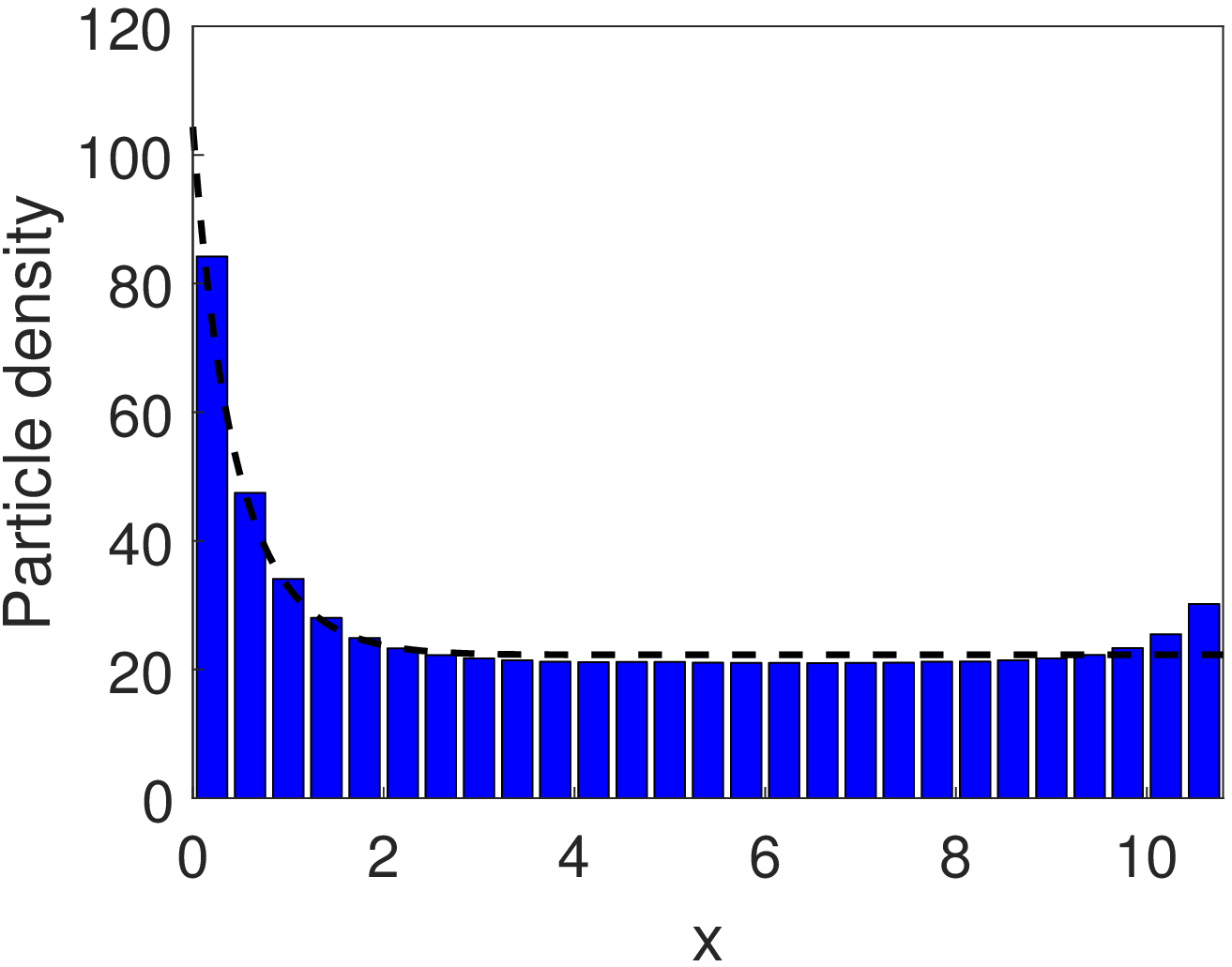}
		\label{fig:Morphogen_orig_tf}}
		\subfigure[][]{
		\includegraphics[width=0.31\textwidth]{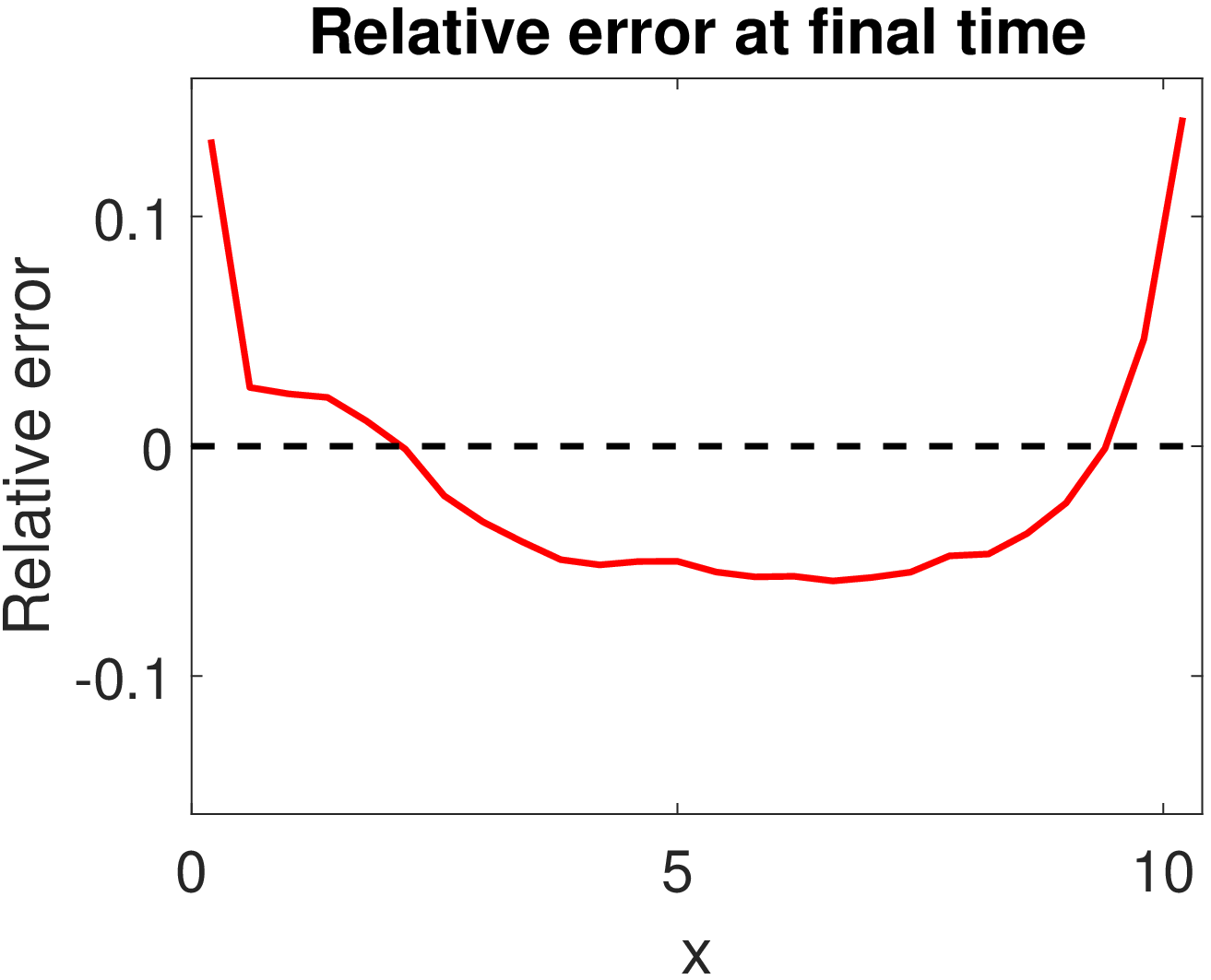}
		\label{fig:Morphogen_orig_rel}}
		\subfigure[][]{
		\includegraphics[width=0.31\textwidth]{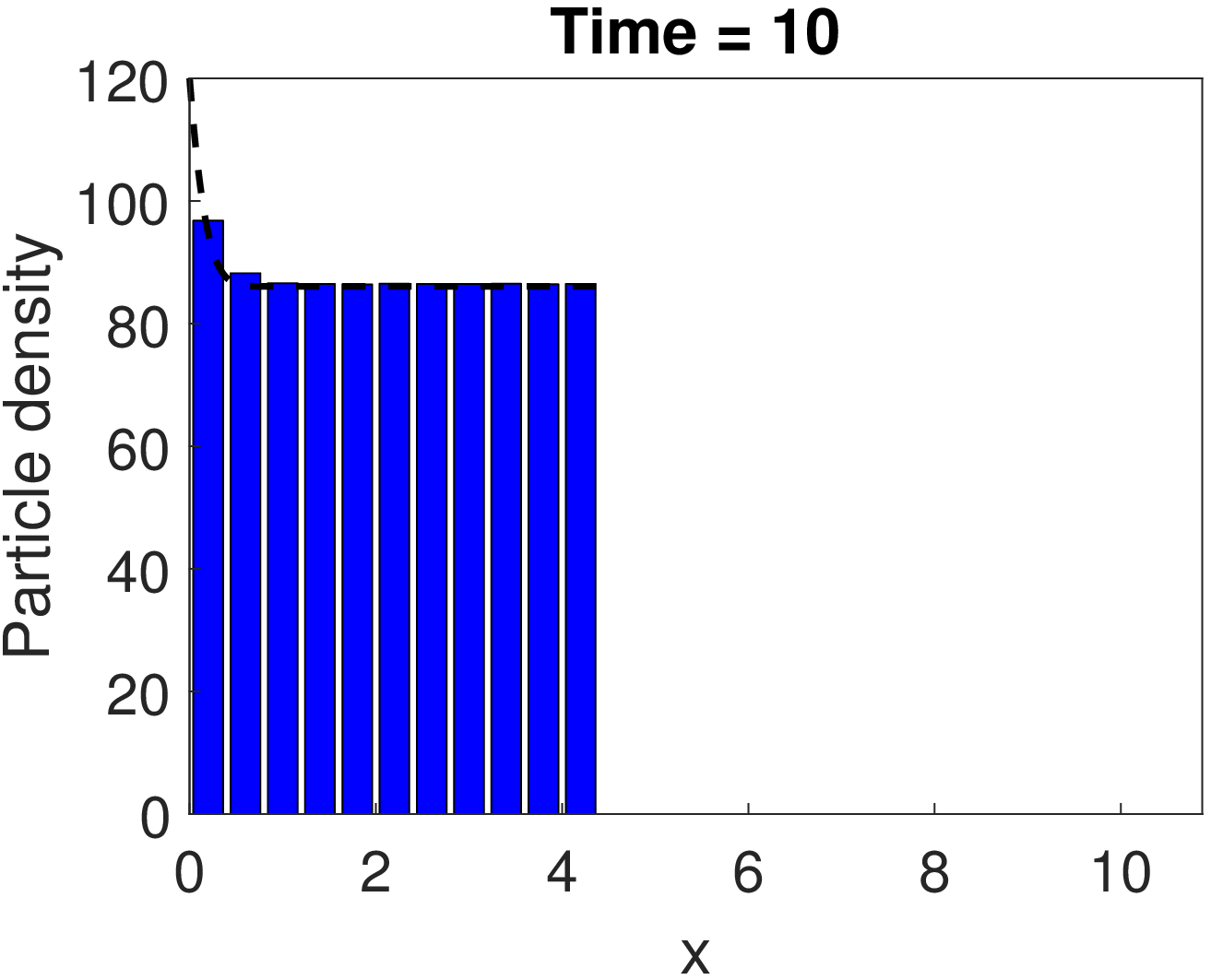}
		\label{fig:Morphogen_stretch_mid}}
		\subfigure[][]{
		\includegraphics[width=0.31\textwidth]{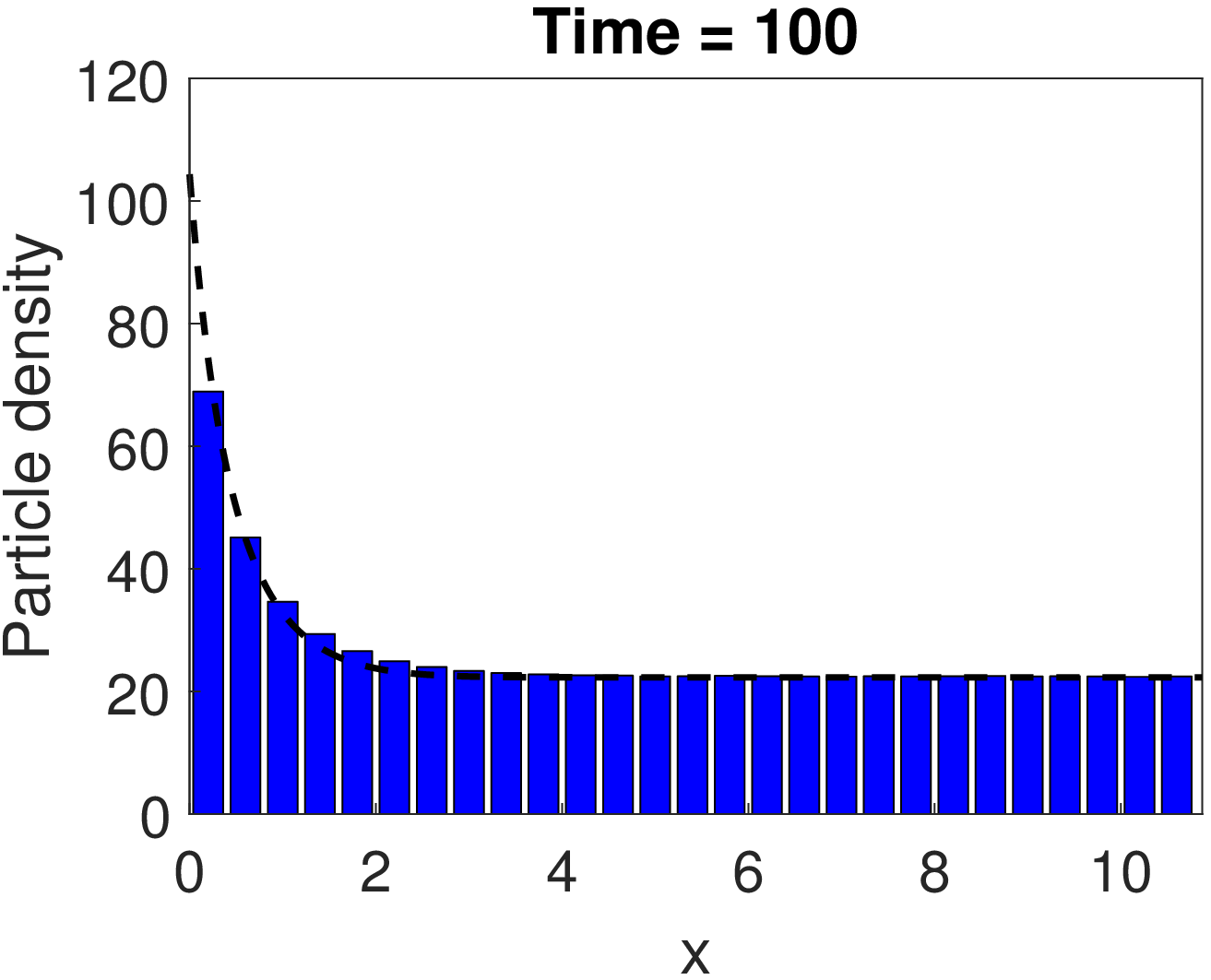}
		\label{fig:Morphogen_stretch_tf}}
		\subfigure[][]{
		\includegraphics[width=0.31\textwidth]{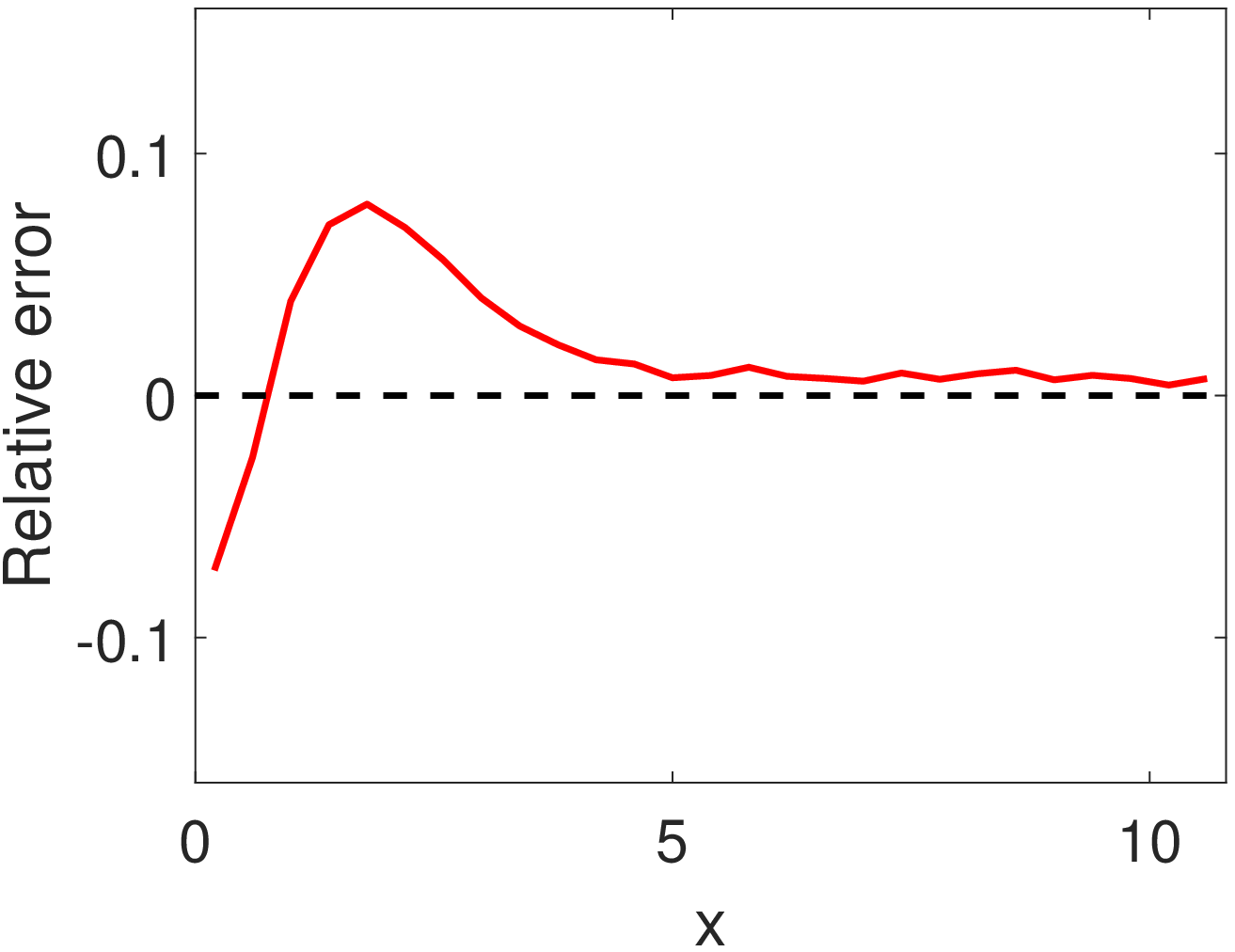}
		\label{fig:Morphogen_stretch_rel}}
		\caption{Two simulations of a morphogen gradient formation using \subref{fig:Morphogen_orig_mid}-\subref{fig:Morphogen_orig_rel} the original method and \subref{fig:Morphogen_stretch_mid}-\subref{fig:Morphogen_stretch_rel} the stretching method. In the first two columns, the blue bars represent the ensemble average of the stochastic algorithm over 50,000 repeats \subref{fig:Morphogen_orig_mid}, \subref{fig:Morphogen_stretch_mid} at time 10 and \subref{fig:Morphogen_orig_tf}, \subref{fig:Morphogen_stretch_tf} at time 100, while the black dashed line is the numerical solution to the PDE \eqref{eqn:Morph_PDE}. In the final column we plot the relative errors between the on-lattice method and the PDE. All parameters are as in the text.}
		\label{fig:Morphogen}
	\end{center}
\end{figure}

We simulate this system with exponential growth rate $\rho=0.01$ on a domain of initial length $L_0=4$. We set the diffusion coefficient to be $D=0.0025$, the influx rate is specified by setting $\lambda = 200$ and the degradation rate $\mu=0.005$. We simulate until a final time, $t=100$ and average over 50,000 repeats. The 
results are displayed in Figure \ref{fig:Morphogen}. 

As with the case of maintaining a uniform gradient, particle densities for the stretching method agree well with the associated mean-field PDE \citep{baker2009fmm} (see Figures \ref{fig:Morphogen_stretch_mid}-\ref{fig:Morphogen_stretch_rel}), however we still observe the same collection of particles at the boundaries for the original method (see Figures \ref{fig:Morphogen_orig_mid}-\ref{fig:Morphogen_orig_rel}). This is particularly evident at the right-hand side of the domain, indicating that we are able to correctly simulate a reaction-diffusion system which incorporates first-order reactions using the stretching method. We anticipate that the extension to second- and higher-order reactions will yield similar results since the domain growth mechanisms is decoupled from the reaction mechanism.


\subsection{Comparison of methods} \label{sect:Stretch_comparison}

In this section, we will investigate the two methods, and the parameter regimes in which the errors from the original method are acceptable due to the interplay between diffusion, the growth rate and initial domain length. We explore this in two ways. The first is through a heuristic argument. 

The mean squared displacement describes how the variance in position of a Brownian particle changes in time. If multiple particles are initialised at the origin and diffuse for a time, $t$, then $\langle x^2 \rangle = 2Dt$, where the angled brackets denote an ensemble average of the squared distances from the origin. If a particle is to diffuse over the entire domain before the domain grows, then the squared distance from the origin would be $x^2 \sim L_0^2$. Likewise, the typical time frame for growth is given by setting $t \sim 1/\rho$. Substituting these into the expression for the mean squared displacement yields $D \sim L_0^2\rho$. Therefore, we say we are in a `high diffusion' parameter regime when $D > L_0^2\rho$. When in the low diffusion regime ($D < L_0^2\rho$), particles are unable to spread and equilibrate before the domain grows, which is when we see the build up of particles in the original method.

\begin{figure}[h!!!!]
	\centering
	\subfigure[][]{
	\includegraphics[width=0.42\textwidth]{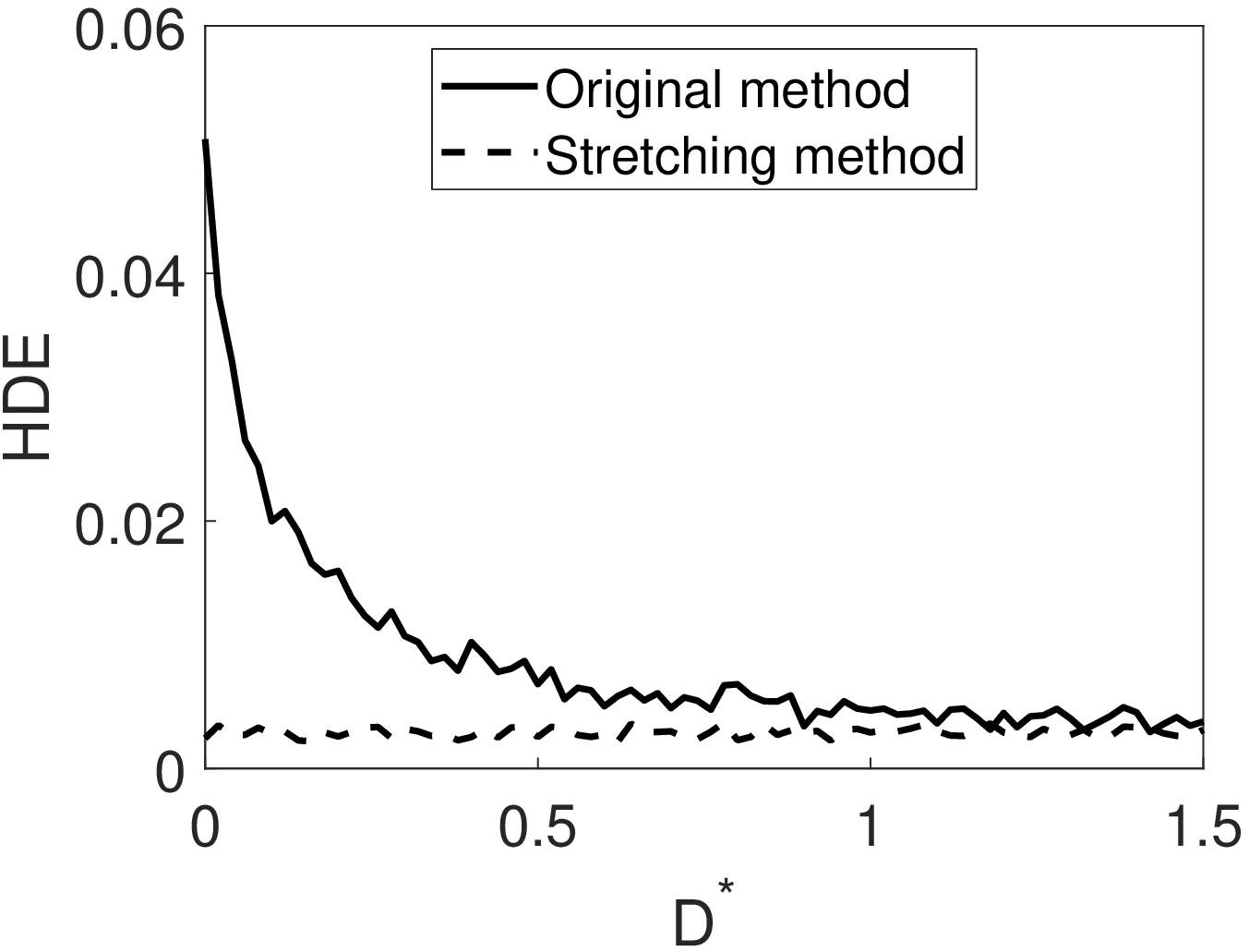}
	\label{fig:Threshold_Stoch}}
	\subfigure[][]{
	\includegraphics[width=0.42\textwidth]{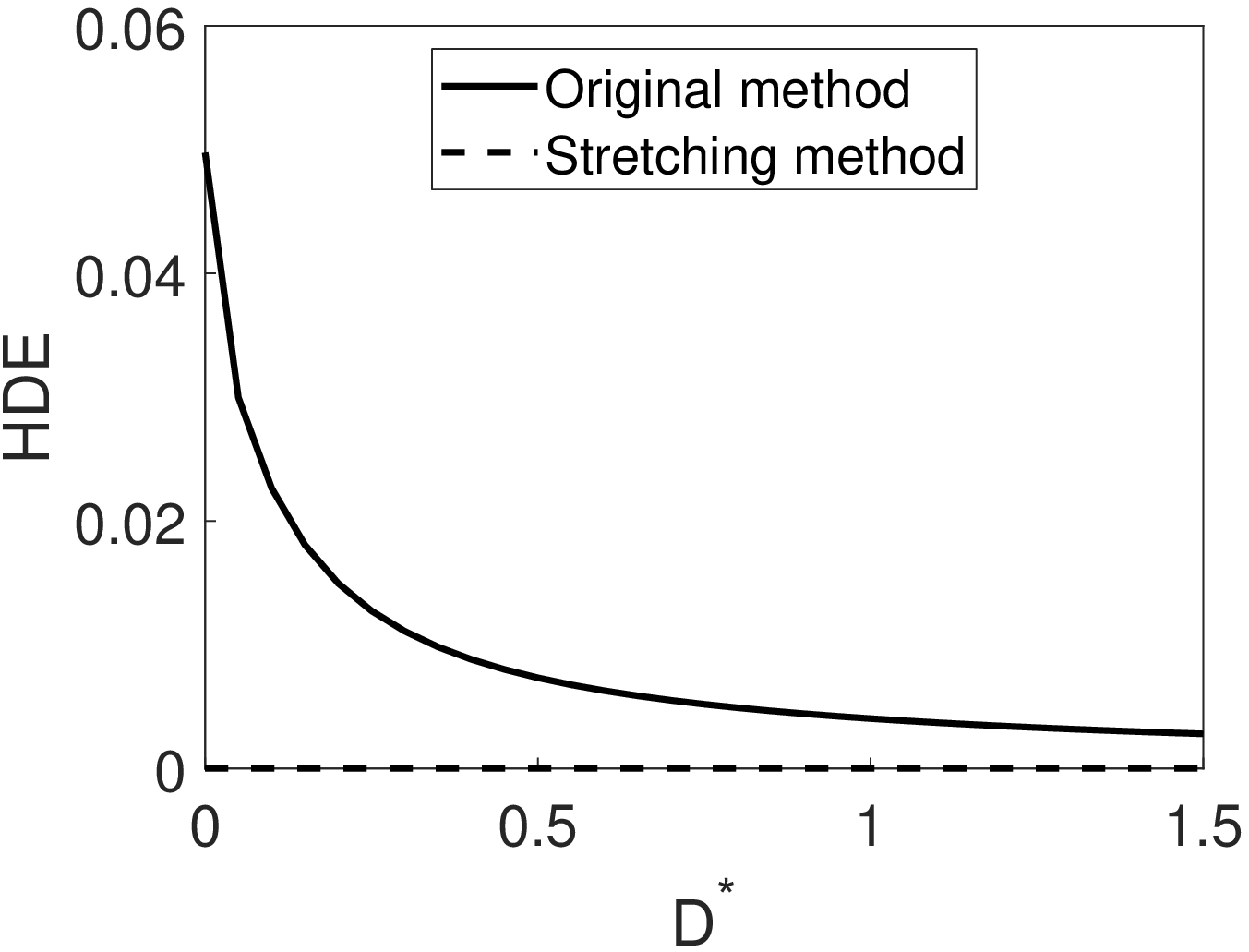}
	\label{fig:Threshold_Mean}}
	\caption{\subref{fig:Threshold_Stoch} HDEs for stochastic simulations of a non-dimensional diffusion mechanism on a growing domain, and their corresponding histogram distance errors (HDEs) for varying diffusion values. The solid line corresponds to the original method \citep{baker2009fmm}, while the dashed line denotes the stretching method. Each simulation is averaged over 5000 repeats. \subref{fig:Threshold_Mean} A similar plot, but using the solutions of the mean equations \eqref{eqn:Growing_MF} and \eqref{eqn:Stretch_Mean_Equations}. Note that the dotted line in this case is on top of the horizontal axis since there is no stochastic error in the mean equations and no bias in the stretching domain growth method.}
	\label{fig:Threshold}
\end{figure}

We verify this heuristic result with a second argument. In order to do this, we have simulated a non-dimensionalised stochastic system and compared it to the solution of the mean-field continuum diffusion equation in order to determine a threshold value for diffusion. The results can be seen in Figure \ref{fig:Threshold}, where we plot the histogram distance error (HDE), which is given by the expression \begin{equation}
\text{HDE}(D^*) = \frac{1}{2}\sum_{n=1}^N{|p_n^s(D^*)-p_n^m(D^*)|}.
\label{eqn:HDE}
\end{equation} Here, $p_n^s(D^*)$ is the value of the normalised solution of the stochastic simulation (original or stretching method) at the final time with non-dimensional diffusion $D^*$, $p_n^m(D^*)$ is the solution for the associated non-dimensional mean-field PDE, and $n$ indexes a common mesh on which we compare the two solutions. The non-dimensional diffusion parameter is equal to $D^*=D/(L_0^2\rho)$ where, as before, $L_0$ is the initial domain length and $\rho$ is the exponential growth rate. From Figure \ref{fig:Threshold_Stoch}, a value $D^*$ greater than 1 yields similar HDE values for both the original and stretching methods. This indicates that, if $D > L_0^2\rho$ then the original method should have a similar performance to the stretching method. However, if the inequality is reversed, so that $D < L_0^2\rho$, then the stretching method should be used. We also note that the same pattern is apparent when comparing the solutions of the mean equations \eqref{eqn:Growing_MF} and \eqref{eqn:Stretch_Mean_Equations}, which can be seen in Figure \ref{fig:Threshold_Mean}. The HDE for the stretching method is exactly zero because it maintains uniformity precisely in the mean field.

\section{Discussion} \label{sect:Discussion}

 
Domain growth mechanisms for on-lattice models are of importance for the accurate representation of many biological processes. We have demonstrated beyond doubt that the original method, suggested by \citet{baker2009fmm}, causes a build up of particles at the boundaries of the spatial domain (see Figure \ref{fig:Growing_meso_toch_prob_1_time_tf}). Consequently, we have developed a method for implementing domain growth when modelling reaction-diffusion systems at the mesoscale in order to correct this build-up. This technique involves stretching all compartments by a small amount (leading to the creation of a new compartment) and the appropriate re-distribution of the particles. We have demonstrated that this method agrees with the corresponding mean-field equations derived in the continuum limit, while maintaining a uniform profile, and have shown that it correctly models morphogen gradient formation on a growing domain.



The stretching method will be particularly useful when developing spatially extended hybrid methods on growing domains. These methods split the spatial domain into subdomains, in which different modelling paradigms are used, separated by an interface or overlap region \citep{smith2018seh}. We envisage that the stretching method can be used in the compartment-based subdomain of a hybrid model for reaction-diffusion on a growing domain, without causing a build-up of particles at the interface.

There are still several open questions regarding on-lattice domain growth whose answers go beyond the scope of this paper. The first of these concerns modelling domain growth in higher dimensions. To induce on-lattice domain growth in higher dimensions we can employ the following method by \citet{binder2008mpt}. Consider, for the purposes of this example, a two-dimensional domain (although higher dimensional growth is straightforward to implemented by analogy). In order to maintain a rectangular domain, a ``growth event'' must increase the total number of rows or columns by one. For example, when a growth event is chosen to occur in the vertical direction, we must increase in the number of rows. To do this, we temporarily treat each column as its own one-dimensional domain, and implement a single vertical growth event in each column, independent of the others. Doing this for every column results in the whole domain increasing in height by a single row. 

\begin{figure}[h!!!!!!!!]
\centering
\includegraphics[width=0.5\textwidth]{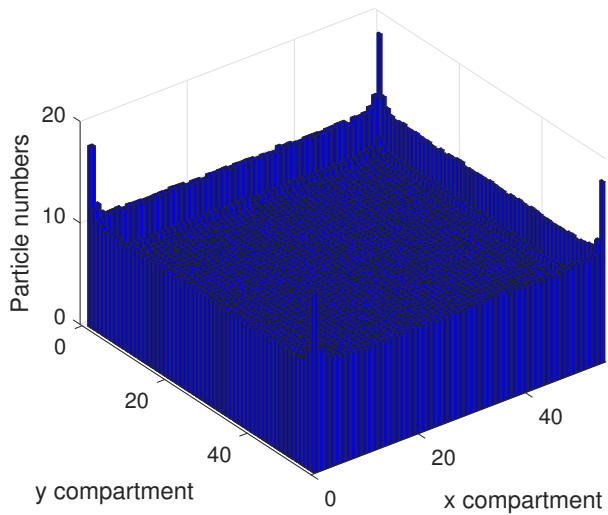}
\caption{A simulation of the domain growth method of \citet{baker2009fmm} extended to two dimensions. The method of domain growth is described in the text. The domain contains 1,000 particles, and particle redistribution is due to domain growth events alone as diffusion is set to zero. The growth rate is $\rho=0.01$, and the particle densities are averaged over 10,000 independent repeats.}
\label{fig:2D_Baker}
\end{figure}

We have simulated such a domain growth process using the original method of \citet{baker2009fmm} to implement the independent row or column elongations when carrying out a horizontal or vertical (respectively) growth event. Specifically, for clarity, we carry out horizontal and vertical growth events simultaneously, which maintains the aspect ratio of the initially square domain we begin with. Diffusion of particles is turned off in order to clearly demonstrate the bias induced by this two-dimensional version of domain growth as illustrated in Figure \ref{fig:2D_Baker}. The same effect that we have observed in one dimension (namely a preponderance of particles towards the boundaries of the domain) is also present in higher dimensions. Extending the stretching method will provide a straightforward fix to this problem in higher dimensions. Overlap fractions will be calculated as ratios of (hyper-)volumes as opposed to ratios of lengths, and particles in a pregrowth compartment will be distributed between multiple overlapping postgrowth compartments using multinomial distributions (the natural generalisation of the binomial distributions used in the one-dimensional case).

Whilst domain growth is important, equally, domain shrinkage is also of significant biological interest. Domain shrinkage through directed apoptosis (programmed cell death) is an important component of many wound healing processes \citep{greenhalgh1998raw, grinnell1999rmt} for example. Without further investigation it is not immediately clear whether domain shrinkage implemented by the removal of a randomly chosen compartment (in analogy to the domain \textit{growth} method of \citet{baker2009fmm} and as implemented by \citet{yates2014dcm}) would induce bias in cell densities. In contrast we are confident that implementing domain shrinkage into the stretching method by considering pre- and post-shrink overlap regions compartments will not induce bias. Nevertheless these hypotheses remain to be tested.

Finally, it is of interest to adapt the stretching method to account for non-uniform growth, which has been shown to be important in biological scenarios \citep{simpson2006cad}. All of the  examples we have presented have implemented uniform growth --- in which all regions of space grow at the same rate. The stretching method could be adapted to incorporate non-uniform growth by splitting the domain into groups of compartments each of which have a different growth rate. The stretching method can then be used on each of the groups individually.

Many authors have used the method presented by \citet{baker2009fmm} to incorporate growth into on-lattice simulations of reaction-diffusion processes. For example, \citet{woolley2011srd} investigate the role that domain growth plays on modelling stochastic reaction-diffusion systems. \citet{thompson2012mcm} explore cell migration and adhesion during biological development, while tissue growth and shrinkage are studied by \citet{yates2014dcm}. However, as demonstrated in this paper, particularly in the case of low diffusivity, the inherent bias in the domain growth method suggests that the conclusions drawn from these studies may require re-evaluation. We suggest that the `stretching' domain growth method we propose in this paper is an appropriate alternative with which to re-evaluate these results and which should be employed in future studies of reaction-diffusion processes on growing domains.

\begin{appendices}

\section{Deriving equation \eqref{eqn:asymp_approx}} \label{sect:appendix_1}

Within this appendix, we derive the expression in equation 
\eqref{eqn:asymp_approx}: 
\begin{equation}
\prod_{n=a}^b{\left(1+\frac{c}{n}\right)} \approx \frac{\Gamma(a)}{\Gamma(a+c)e^c}b^c, \tag{\ref{eqn:asymp_approx}}
\end{equation}
which holds for $a,b\in\mathbb{N}$ such that $a<b$ and $b$ is large, and for $c\in\mathbb{R}$. We start by representing the product on the left-hand side of equation \eqref{eqn:asymp_approx} using gamma functions: \begin{align}
\prod_{n=a}^b{\left(1+\frac{c}{n}\right)} &= \prod_{n=a}^b{\left(\frac{n+c}{n}\right)}\notag\\
&= \frac{\Gamma(a)\Gamma(b+c+1)}{\Gamma(a+c)\Gamma(b+1)}. \tag{A1}\label{eqn:App_1_Int_1}
\end{align}
We apply Stirling's approximation, which says that for large z, $\Gamma(z)\approx\sqrt{2\pi(z-1)}(z-1)^{z-1}e^{-(z-1)}$. Applying this to equation \eqref{eqn:App_1_Int_1}, we obtain: \begin{align*}
\prod_{n=a}^b{\left(1+\frac{c}{n}\right)} &\approx \frac{\Gamma(a)}{\Gamma(a+c)}\frac{\sqrt{2\pi(b+c)}}{\sqrt{2\pi b}}\frac{(b+c)^{b+c}}{b^b}\frac{e^b}{e^{b+c}}\\
&= \frac{\Gamma(a)}{\Gamma(a+c)}\sqrt{\frac{b+c}{b}}\left(\frac{b+c}{b}\right)^b(b+c)^ce^{-c}.
\end{align*}
We assume that $b$ is large, and thus, $b+c\approx b$. Applying this: \begin{align*}
\prod_{n=a}^b{\left(1+\frac{c}{n}\right)} &\approx \frac{\Gamma(a)}{\Gamma(a+c)}\sqrt{\frac{b}{b}}\left(\frac{b}{b}\right)^bb^ce^{-c}\\
&= \frac{\Gamma(a)}{\Gamma(a+c)e^c}b^c.
\end{align*} 

\section{Deriving equation \eqref{eqn:terms2_simplified}} 
\label{sect:appendix_2}

Within this appendix, we derive the expression in equation 
\eqref{eqn:terms2_simplified}: \begin{equation}
\sum_{n=2}^{k-1}{\left(\prod_{j=n+1}^k{\left(1-\frac{3/2}{j}\right)}
\right)M_1^n\left(\frac{1/2}{n}\right)} \approx \frac{1}{2}\frac{1}{\sqrt{\pi}e^{-1/2}}k^{-1/2} - \frac{1}{\sqrt{\pi}e^{-1/2}}k^{-3/2},
\tag{\ref{eqn:terms2_simplified}}
\end{equation}
where the approximation is for large values of $k$.
\begin{align*}
\sum_{n=2}^{k-1}{\left(\prod_{j=n+1}^k{\left(1-\frac{3/2}{j}\right)}\right)M_1^n\left(\frac{1/2}{n}\right)} &= \sum_{n=2}^{k-1}{\left(\prod_{j=1}^k{\left(1-\frac{3/2}{j}\right)}\right)\left(\prod_{j=1}^n{\left(1-\frac{3/2}{j}\right)}\right)^{-1}M_1^n\left(\frac{1/2}{n}\right)}\\
&= \left(\prod_{j=1}^k{\left(1-\frac{3/2}{j}\right)}\right)\sum_{n=2}^{k-1}{\left(\prod_{j=1}^n{\left(1-\frac{3/2}{j}\right)}\right)^{-1}M_1^n\left(\frac{1/2}{n}\right)}\\
&\approx k^{-3/2}\sum_{n=2}^{k-1}{\left(n^{-3/2}\right)^{-1}M_1^n\left(\frac{1/2}{n}\right)},
\end{align*}
where, in the final approximation, we have used equation \eqref{eqn:asymp_approx} in order to simplify the two products. Simplifying yields:
\begin{equation*}
\sum_{n=2}^{k-1}{\left(\prod_{j=n+1}^k{\left(1-\frac{3/2}{j}\right)}\right)M_1^n\left(\frac{1/2}{n}\right)} \approx k^{-3/2}\sum_{n=2}^{k-1}{n^{3/2}M_1^n\left(\frac{1/2}{n}\right)}.
\end{equation*}
We now utilise equation \eqref{eqn:asymp_n1_sol} (approximation is better for large $k$), in order to obtain:
\begin{align*}
\sum_{n=2}^{k-1}{\left(\prod_{j=n+1}^k{\left(1-\frac{3/2}{j}\right)}\right)M_1^n\left(\frac{1/2}{n}\right)} &\approx k^{-3/2}\sum_{n=2}^{k-1}{n^{3/2}\frac{1}{\sqrt{\pi}e^{-1/2}}n^{-1/2}\frac{1}{2}n^{-1}}\\
&= k^{-3/2}\frac{1}{\sqrt{\pi}e^{-1/2}}\sum_{n=2}^{k-1}{\frac{1}{2}}\\
&= k^{-3/2}\frac{1}{\sqrt{\pi}e^{-1/2}}\frac{1}{2}\left(k-2\right)\\
&= \frac{1}{2}\frac{1}{\sqrt{\pi}e^{-1/2}}k^{-1/2} - \frac{1}{\sqrt{\pi}e^{-1/2}}k^{-3/2}.\\
\end{align*}

\section{Discussion on local methods} \label{sect:Appendix_Local}
Throughout Section \ref{sect:Orig}, we have demonstrated that at low diffusion levels, the method presented by \citet{baker2009fmm} fails to correctly model the growth of a uniform profile. This is a local method, which means that on every occasion that the domain is due to grow, we choose a compartment at random at which the growth event occurs. Over many repeats of the same process, different compartments will be chosen and so, when averaging over these repeats, each compartment is chosen an equal number of times (when considering only a single growth event).

There are three elements that define a local method: \begin{enumerate}
\item The probability of choosing each compartment,
\item The direction of growth,
\item The redistribution of particles.
\end{enumerate} In the case of \citet{baker2009fmm}, we (1) choose each compartment uniformly at random, (2) always grow to the right and (3) redistribute the particles using a symmetric binomial distribution. We now discuss which of these we can change in order to create a local method that correctly maintains a uniform profile on a growing domain.

The simplest way of creating a different local method is to change one of the three elements. However, it can be shown that changing only a single element does not yield the expected uniform growth. As a result, the next simplest is to change two of the elements, whilst fixing one. One such way would be to (1) set the probability of choosing each compartment to be general, (2) set the direction of growth to be left or right with a probability of a half each, and (3) redistribute the particles using a binomial distribution with a generic probability of success. These properties yield an algorithm which has $k$ degrees of freedom (where there are $k-1$ pregrowth compartments), which can be used in order to solve a series of equations to ensure that a uniform profile is maintained. 

In order to calculate the values for the probability distribution, and the probability for success in the binomial distribution when redistributing particles post-split, we can write a series of equations that relate the pre- and postgrowth states, on average. Using these, we are able to use numerical optimisation techniques in order to find the optimum values for the unknown parameters. However, while this method fixes the main issue with the original method, it introduces some new ones. Firstly, the centre of growth is no longer fixed at one end of the domain. This may cause problems for certain methodological applications (e.g. employing the method for spatial hybrid methods \citep{yates2015pcm,flegg2015cmc,smith2017arm}) or for some biological applications in which tissues genuinely grow from a fixed origin at one end of the domain grow from one end of the domain (e.g. hyphal tip growth \citep{goriely2008mmh}). Secondly, we have to solve an overdetermined system for every possible number of compartments that might occur. This can be computationally expensive, especially for large compartment numbers. In practice, the unknown parameters can be computed and stored \textit{a priori}, although, if the timing of domain growth events is stochastic, it may not be clear in advance exactly how many compartments the domain will comprise.

We also note that we have changed the direction of growth from being always to the right in the original method, to being left or right with equal probability. This is because there is no probability distribution for the compartments together with redistribution probability that maintains a uniform profile when the direction of growth is always the same.

\section{Justifying the stretching method} \label{sect:Appendix_Stretch}

In Section \ref{sect:Stretch}, we introduced the stretching method as an alternative to the original method presented by \citet{baker2009fmm}. In order to demonstrate that the original method fails, we have simulated the corresponding mean equations, and also analytically calculated the density of particles in each compartment with a large number of compartments. In this section, we will conduct a similar analysis of the stretching method.

\subsection{The mean equations} \label{sect:Appendix_Stretch_Mean}

In this section, we will calculate the mean equations for the stretching method by firstly considering the master equation for the process. We will then simulate the solutions to the mean equations, demonstrating that the solution remains uniform.

We begin by defining the sets $\mathcal{M}_k^N = \set{\bs{m}\in\mathbb{N}^k}{\sum_{i=1}^k{m_i}=N}$. This is the set of all state vectors when there are $k$ compartments and $N$ particles in the system in total. We would like an expression for the probability that there are $k$ compartments in total, and the state variable is $\bs{m}\in\mathcal{M}_k^N$ with $N$ a fixed integer. We call this probability $p(\bs{m},k,t)$ as we did in Section \ref{sect:Orig}. Then, defining $\pi(\bs{m}|\bs{r})$ to be the transition probability from state $\bs{r}$ to state $\bs{m}$: \begin{equation}
\ordder{p}{t}(\bs{m},k,t) =  \rho(k-1)\sum_{\bs{r}\in\mathcal{M}_{k-1}^N}{\left[p(\bs{r},k-1,t)\pi\left(\bs{m}|\bs{r}\right)\right]}-\rho kp(\bs{m},k,t).
\label{eqn:Stretch_ME}
\end{equation} Here, the summand in the first term of the right hand side represents the rate of moving from a state $\bs{r}\in\mathcal{M}_{k-1}^N$ to the state $\bs{m}\in\mathcal{M}_k^N$. The second term is the rate at which the process leaves the state $\bs{m}$.

We next derive the mean equations. Similarly to the original method, define: \begin{equation}
M_i^k(t) = \sum_{\bs{m}\in\mathcal{M}_k^N}{m_ip(\bs{m},k,t)}.
\label{eqn:Stretch_Mean}
\end{equation} Multiplying the CME \eqref{eqn:Stretch_ME} by $m_i$, summing over all possible $\bs{m}\in\mathcal{M}_k^N$, and applying equation \eqref{eqn:Stretch_Mean}, we obtain: \begin{equation}
\ordder{M_i^k}{t}(t) = \rho(k-1)\sum_{\bs{m}\in\mathcal{M}_k^N}{\sum_{\bs{r}\in\mathcal{M}_{k-1}^N}{\left[m_ip(\bs{r},k-1,t)\pi\left(\bs{m}|\bs{r}\right)\right]}} - \rho kM_i^k(t).
\label{eqn:Stretch_Mean_Eqs_Int}
\end{equation} From now on, we will drop the range of the sums, and simply write $\bs{m}$ or $\bs{r}$ and implicitly assume we are summing over the correct sets in order to simplify the notation. Recall that for each $\bs{r}$, there is an associated vector of binomial random variables $\bs{b}$ that re-distributes the particles from the pregrowth state to the postgrowth state. Further, for any $j\in\{1,...,k-1\}$, the probability of drawing $b_j$ is given by the probability mass function: \begin{equation}
\mathbb{P}(b_j|\bs{r}) = {r_j \choose b_j}\left(\delta_j^{k-1}\right)^{b_j}(1-\delta_j^{k-1})^{r_j-b_j},
\label{eqn:Stretch_pmf}
\end{equation} where $\delta_j^{k-1}$ is the `overlap' proportion defined in equation \eqref{eqn:Overlap_Proportion}. We can then use the relationships set out in Algorithm \ref{alg:Stretch}, namely: \begin{equation}
m_j = \left\{\begin{array}{ll}
	b_1, & \text{if }j = 1, \\ 
	b_j + (r_{j-1}-b_{j-1}), & \text{if }j\in\{2,...,k-1\}, \\ 
	r_{k-1}-b_{k-1}, & \text{if }j = k.
	\end{array} 
	\right.
\label{eqn:Stretch_Redist}
\end{equation} Assuming that $i \in \{2,...,k-1\}$ (a similar argument can be applied to the case $i\in\{1,k\}$) and substituting the relationships in \eqref{eqn:Stretch_Redist} into the sum in equation \eqref{eqn:Stretch_Mean_Eqs_Int}, we obtain: \begin{align}
\sum_{\bs{m}}{\sum_{\bs{r}}}&{{m_ip(\bs{r},k - 1,t)\pi\left(\bs{m}|\bs{r}\right)}} = \sum_{\bs{m}}{\sum_{\bs{r}}{(r_{i-1}-b_{i-1}+b_i)p(\bs{r},k-1,t)\pi\left(\bs{m}|\bs{r}\right)}}\notag\\
&= \sum_{\bs{r}}{\sum_{\bs{m}}{r_{i-1}p(\bs{r},k-1,t)\pi\left(\bs{m}|\bs{r}\right)}} - \sum_{\bs{r}}{\sum_{\bs{m}}{(b_{i-1}-b_i)p(\bs{r},k-1,t)\pi\left(\bs{m}|\bs{r}\right)}},\label{eqn:Stretch_Mean_Eqs_Split}
\end{align} where in the final step, we have split the sum and also switched the order of summation. We will now make use of the binomial random variables (the $b_j$'s). We note that in order to transition from the pregrowth state $\bs{r}\in\mathcal{M}_{k-1}^N$ to the postgrowth state $\bs{m}\in\mathcal{M}_k^N$, we need to find a vector $\bs{b}$. However, only a small subset of these $\bs{b}$ vectors have a non-zero probability of occurring. Therefore, the sum over $\bs{m}$ can be re-written as a sum over the possible $\bs{b}$ values that have a non-zero probability. Using this, and substituting for $\pi(\bs{m}|\bs{r})$ the specific probability $\mathbb{P}(\bs{b}|\bs{r})$ of the binomial redistribution which takes us from state $\bs{r}$ to state $\bs{m}$, equation \eqref{eqn:Stretch_Mean_Eqs_Split} becomes: \begin{align}
\sum_{\bs{r}}{\sum_{\bs{m}}{m_ip(\bs{r},k-1,t)\pi\left(\bs{m}|\bs{r}\right)}} &= \sum_{\bs{r}}{\sum_{\bs{m}}{r_{i-1}p(\bs{r},k-1,t)\pi\left(\bs{m}|\bs{r}\right)}} - \sum_{\bs{r}}{\sum_{\bs{m}}{(b_{i-1}-b_i)p(\bs{r},k-1,t)\pi\left(\bs{m}|\bs{r}\right)}}\notag\\
&= \sum_{\bs{r}}{\sum_{\bs{b}}{r_{i-1}p(\bs{r},k-1,t)\mathbb{P}\left(\bs{b}|\bs{r}\right)}} - \sum_{\bs{r}}{\sum_{\bs{b}}{(b_{i-1}-b_i)p(\bs{r},k-1,t)\mathbb{P}\left(\bs{b}|\bs{r}\right)}}. \label{eqn:Mean_Eqs_Int}
\end{align} Here $\mathbb{P}(\bs{b}|\bs{r}) = \prod_{j=1}^{k-1}{\mathbb{P}(b_j|\bs{r})}$ (where $\mathbb{P}(b_j|r_j)$ are given in equation \eqref{eqn:Stretch_pmf}), and the sum over $\bs{b}$ is a sum over the set $\set{\bs{b}}{b_j\in\{0,...,r_j\}\ \forall j\in\{1,...,k-1\}}.$ We now re-arrange the two sums in expression \eqref{eqn:Mean_Eqs_Int}: \begin{align*}
\sum_{\bs{r}}{\sum_{\bs{b}}}&{{r_{i-1}p(\bs{r},k-1,t)\mathbb{P}\left(\bs{b}|\bs{r}\right)}} - \sum_{\bs{r}}{\sum_{\bs{b}}{(b_{i-1}-b_i)p(\bs{r},k-1,t)\mathbb{P}\left(\bs{b}|\bs{r}\right)}}\\
&= \sum_{\bs{r}}{r_{i-1}p(\bs{r},k-1,t)\left[\sum_{\bs{b}}{\mathbb{P}(\bs{b}|\bs{r})}\right]} - \sum_{\bs{r}}{p(\bs{r},k-1,t)\left[\sum_{\bs{b}}{(b_{i-1}-b_i)\mathbb{P}(\bs{b}|\bs{r})}\right]}.
\end{align*} The sum in the square brackets in the first term is equal to 1 as it is the sum of a probability distribution and the sum in the square brackets in the second term is a difference of two expectations: \begin{align*}
\sum_{\bs{r}}{r_{i-1}}&{p(\bs{r},k-1,t)\left[\sum_{\bs{b}}{\mathbb{P}(\bs{b}|\bs{r})}\right]} - \sum_{\bs{r}}{p(\bs{r},k-1,t)\left[\sum_{\bs{b}}{(b_{i-1}-b_i)\mathbb{P}(\bs{b}|\bs{r})}\right]}\\
&= \sum_{\bs{r}}{r_{i-1}p(\bs{r},k-1,t)} - \sum_{\bs{r}}{p(\bs{r},k-1,t)\left[\mathbb{E}[b_{i-1}|\bs{r}]-\mathbb{E}[b_i|\bs{r}]\right]}.
\end{align*} Finally, using the standard expectation for the binomial distribution, and bringing terms together, we find that: \begin{align}
\sum_{\bs{m}}{\sum_{\bs{r}}{m_ip(\bs{r}},k-1,t)\pi\left(\bs{m}|\bs{r}\right)} &= (1-\delta_{i-1}^{k-1})\sum_{\bs{r}}{r_{i-1}p(\bs{r},k-1,t)} + \delta_i^{k-1}\sum_{\bs{r}}{r_ip(\bs{r},k-1,t)} \notag\\
&= (1-\delta_{i-1}^{k-1})M_{i-1}^{k-1} + \delta_i^{k-1}M_i^{k-1}, \label{eqn:Stretch_Mean_Eqs_Sum_Simplify}
\end{align} where the final equality uses the definition \eqref{eqn:Stretch_Mean}. Substituting this expression into equation \eqref{eqn:Stretch_Mean_Eqs_Int} yields the mean-field density evolution equations for the stretching method: \begin{equation}
\ordder{M_i^k}{t} = \rho(k-1)\left[(1-\delta_{i-1}^{k-1})M_{i-1}^{k-1} + \delta_i^{k-1}M_i^{k-1}\right] - \rho kM_i^k. \label{eqn:Stretch_Mean_Equations}
\end{equation}

\begin{figure}[h!!!!!!!!!]
\centering
\includegraphics[width=0.5\textwidth]{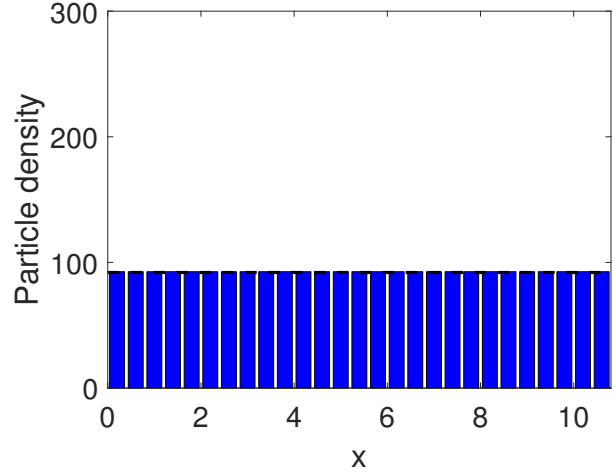}
\caption{The solution to the mean-field equations \eqref{eqn:Stretch_Mean_Equations}. All parameters are the same as Figure \ref{fig:Growing_mean_field_Eqs}.}
\label{eqn:Mean_field_Stretch}
\end{figure}

\subsection{Analytical density} \label{sect:Appendix_Stretch_Analytical}

Here we will use a similar approach to Section \ref{sect:Stretch} in order to calculate the average density of particles when the domain grows according to the stretching method. We let $u_i^k$ be the random density of particles in compartment $i$ when there are $k$ compartments in total, $M_i^k = \mathbb{E}[u_i^k]$, and $q(u_i^k,\delta_i^k)$ a realisation from a general probability distribution with mean value $M_i^k\delta_i^k$. The $q(u_i^k,\delta_i^k)$ values denote the density that is redistributed into postgrowth compartment $i$ from pregrowth compartment $i$, analogous to the $\mathbb{P}(b_i|r_i)$ in equation \eqref{eqn:Stretch_pmf} for the discretised process. Then the $u_i^k$'s are related by the following recursive relation: \begin{equation}
u_i^k = q\left(u_i^{k-1},\delta_i^{k-1}\right) + \left(u_{i-1}^{k-1} - q\left(u_{i-1}^{k-1},\delta_{i-1}^{k-1}\right)\right)
\label{eqn:Stretch_Recursion_Random}
\end{equation} with an initial condition $u_1^1 = 1$. Here, the first term on the right-hand side is the fraction of the density that is provided to compartment $i$ on the postgrowth domain by the pregrowth compartment to the right, while the second term is the density provided from the left pregrowth compartment. We now take expectations to yield the recursive equations: \begin{equation}
M_i^k = \delta_i^{k-1}M_i^{k-1} + (1-\delta_{i-1}^{k-1})M_{i-1}^{k-1}.
\label{eqn:Stretch_Recursion}
\end{equation} We show that, under this recursion relation, $M_i^k=1/k$ by induction on the number of compartments, $k$. Clearly, using the initial condition, we have the base case $M_1^1=1$. Now assume that $M_i^{k-1} = 1/(k-1)$ for all $i\in\{1,...,k-1\}$. Then, using \eqref{eqn:Stretch_Recursion}: \begin{align*}
M_i^k &= \delta_i^{k-1}M_i^{k-1} + (1-\delta_{i-1}^{k-1})M_{i-1}^{k-1}\\
&= \frac{\delta_i^{k-1}}{k-1} + \frac{1-\delta_{i-1}^{k-1}}{k-1}\\
&= \frac{k-i + (k-(k-i-1))}{k(k-1)}\\
&= \frac{1}{k}.
\end{align*} Here, the second line uses the inductive hypothesis and the third line employs the definition of $\delta_i^{k-1}$. This completes the inductive step, and hence the proof that $M_i^k = 1/k$.

\end{appendices}

\bibliographystyle{unsrtnat}
\bibliography{mesoscale_growth}

\end{document}